\documentclass[a4paper, 12pt]{article}

\usepackage{amsrefs, amsmath}
\usepackage{amsfonts, mathrsfs}
\usepackage{amssymb}
\usepackage{amscd}
\usepackage{fullpage}
\usepackage[labelformat=empty]{subfig}
\usepackage{booktabs}
\usepackage{xypic}
\usepackage[dvips]{graphicx,epsfig}
\usepackage{color}

\newtheorem{thm}{Theorem}[section]

\newcommand{\de}{{\partial}}

\newcommand{\bbN}{\mathbb{N}}

\newcommand{\bbL}{\mathbb{L}}
\newcommand{\bbZ}{\mathbb{Z}}

\newcommand{\bbR}{\mathbb{R}}
\newcommand{\bbC}{\mathbb{C}}
\newcommand{\bbP}{\mathbb{P}}
\newcommand{\bbF}{\mathbb{F}}
\newcommand{\bbQ}{\mathbb{Q}}
\newcommand{\bbT}{\mathbb{T}}

\def\bary{\begin{array}} 
\def\eary{\end{array}} 
\def\ben{\begin{enumerate}} 
\def\een{\end{enumerate}}
\def\bit{\begin{itemize}} 
\def\eit{\end{itemize}}
\def\nn{\nonumber} 

\newcommand{\cO}{\mathcal{O}}

\newcommand{\cE}{\mathcal{E}}

\newcommand{\LL}{\mathcal{L}}

\newcommand{\cN}{\mathcal{N}}

\newcommand{\cA}{\mathcal{A}}

\newcommand{\cF}{\mathcal{F}}
\newcommand{\cI}{\mathcal{I}}

\newcommand{\cX}{\mathcal{X}}

\newcommand{\cM}{\mathcal M}

\newcommand{\Li}{\operatorname{Li}}

\def\beq{\begin{equation}}                     %  
\def\eeq{\end{equation}}                       % 
\def\bea{\begin{eqnarray}}                     %         % 
\def\eea{\end{eqnarray}}

\def\a{\alpha}
\def\b{\beta}

\def\Res{\mathrm{Res}}

\newcommand{\GIT}[1]{/\!\!/_{\kern-.2em #1 \kern0.1em}}

\renewcommand{\l}{\left}
\renewcommand{\r}{\right}
\newcommand{\bra}{\left\langle}
\newcommand{\ket}{\right\rangle}

\newcommand{\re}{\mathrm{e}}
\newcommand{\rd}{\mathrm{d}}

\numberwithin{equation}{section}
\begin{document}

\setcounter{page}{0}
\begin{titlepage}
\titlepage
%\rightline{SISSA-??-2008-FM}
\vskip 3cm
\centerline{{ \bf \Large Open topological strings and integrable    hierarchies:}}
\vskip .5cm
\centerline{{ \bf \Large
    Remodeling the A-model}}
\vskip 1.5truecm
\centering
{\bf Andrea Brini}
\vskip 1.5cm
\begin{center}
\em 
Section de Math\'ematiques and D\'epartement de Physique Th\'eorique \\
Universit\'e de Gen\`eve \\ 
24 quai Ansermet, CH-1211, Geneva, Switzerland

\vskip 3cm
\end{center}
\begin{abstract}
We set up, purely in A-model terms, a novel formalism for the global solution of the open and closed
topological A-model on toric Calabi-Yau threefolds. The starting point is to build on
recent progress in the mathematical theory of open Gromov-Witten invariants of
orbifolds; we interpret the localization formulae as relating D-brane amplitudes to closed
string amplitudes perturbed with twisted masses through an analogue of the ``loop
insertion operator'' of matrix models. We first generalize this form of
open/closed string duality to general toric backgrounds in all chambers of the
stringy K\"ahler moduli space; secondly, we display a neat
connection of the (gauged) closed string side to tau
functions of 1+1 Hamiltonian integrable hierarchies, and exploit it to provide an effective
computation of open string amplitudes. In doing so, we also provide a
systematic treatment of the change of flat open moduli induced by a phase
transition in the closed moduli space. We test our proposal in detail by
providing an extensive number of checks.
%, including amplitudes with non-trivial
%quasi-modular properties
We also use our formalism 
%to provide an A-model
%derivation of known results of local mirror symmetry; in particular,
to give 
a localization-based derivation of the Hori-Vafa spectral curves as coming from a resummation of A-model disc instantons.

\end{abstract}

\vskip 5cm

\vfill
 \hrule width 5.cm
\vskip 2.mm
{\small 
\noindent }
\begin{flushleft}
%\vspace{.5cm}
  \texttt{andrea.brini  at  unige.ch}
\end{flushleft}
\end{titlepage}

\tableofcontents

\section{Introduction}

The topological phase of string theory  has been a major source of insights
both in Mathematics and Physics. Topological strings on Calabi-Yau manifolds
come in two guises, which are related by mirror symmetry: the
A-model and the B-model. On the Physics side, they yield a great deal
of non-trivial information about the vacuum structure of type IIA and IIB superstring
compactifications and the holomorphic effective dynamics of the resulting
supersymmetric gauge and gravity theories;
%\cite{Bershadsky:1993cx, Antoniadis:1993ze, Vafa:2000wi, Dijkgraaf:2002fc,
%Katz:1996fh}; 
at the same time, they provide a
privileged laboratory for studying general ideas about dualities in string
theory, such as mirror symmetry and gauge/string duality. From a mathematical
point of view, the  A-twisted topological string captures a
sophisticated set of invariants of the target manifold in the form of a
``virtual'' count of holomorphic maps in the form of Gromov-Witten invariants;  its  B-model mirror symmetric
counterpart can instead be regarded as a quantized version of the theory of
variation %of deformations 
of Hodge structures on the mirror Calabi-Yau.  \\

A special case, and one that has been subject to intense study in the last
decade, is given by the topological A-model on {\it toric} Calabi-Yau
threefolds in presence of Lagrangian D-branes. From a technical point of view, this setup enjoys a host of
desirable features: it provides a local model for the topological string on
compact Calabi-Yau threefolds, and at the same time it shares many qualitative
features - like the existence of matrix model duals \cite{Marino:2002fk, Aganagic:2002wv,
  Eynard:2010dh}, a mirror description in
terms of Riemann surfaces \cite{Hori:2000kt, Hori:2000ck}, and some form of
underlying integrability \cite{Aganagic:2003qj, Eynard:2007kz, Eynard:2008he} - with the case of
topological strings on Fano manifolds. \\
Essentially two, somewhat complementary formalisms
have been put forward to solve the open and closed topological A-model on
these backgrounds: the topological vertex formalism of \cite{Aganagic:2003qj} and the
``Remodeling the B-model'' proposal of \cite{Bouchard:2007ys}. In both cases, the
key principle at the base is some string duality: for the topological vertex,
topological gauge/string duality with Chern-Simons theory \cite{Gopakumar:1998ki}
allows to solve the theory to all orders in $g_s$ around  large radius; for the
remodeled-B-model, a local mirror symmetry picture in terms of dual spectral curves
is the starting point for a recursive solution based on the Eynard-Orantin
formalism for matrix models \cite{Marino:2006hs,Bouchard:2007ys,Eynard:2007kz}.  \\

In this paper we develop a formalism to solve the open topological A-model on
a toric CY3 target $X$ with toric Lagrangian branes $L\hookrightarrow X$ from
a direct A-model instanton analysis, 
%around
%orbifold points,  
without appealing to string
duality. In our setup, genus $g$, $h$-holed open string amplitudes are computed from (equivariant)
closed string amplitudes via an analogue of the loop insertion operator of
matrix models:
\beq
F^{X,L}_{g,h}(t_1, \dots, t_n; w_1, \dots w_h;f) = \l(\prod_{i=1}^h \LL^X(w_i,f)\r)
\cF^X_{g}(t_{\a,p}, f)\Bigg|_{t_{\a,p}=\delta_{{\rm deg\phi_\a},2 }
  \delta_{p,0}t_\a}.
\label{eq:masterformulaintro}
\eeq
In the right hand side, $\cF^X_{g}(t_{\a,p})$ is the full-descendant genus
$g$ free energy of the topological A-model, twisted by a $T\simeq \bbC^*$
action specified by the location of the brane and the choice of an integer
$f$ which captures the framing ambiguity on the dual Chern-Simons side. The
chemical potentials $t_{\a,p}$ count the insertions of the $p^{\rm th}$
gravitational descendant of a chiral operator $\phi_\a \in QH^\bullet_T(X)$, whereas
the operator $\LL^X(w,f)$ is a first order differential operator in
$t_{\a,p}$. \\

Eq. \eqref{eq:masterformulaintro} is the outcome of the localization approach
to define and compute open Gromov-Witten invariants on toric Calabi-Yau
manifolds first put forward in \cite{kl:open}; in the case of orbifolds of
$\bbC^3$, Atiyah-Bott localization \cite{Brini:2010sw} results in an explicit expression for the
operator $\LL(w,f)$. We will build on this along two main directions. We first
of all generalize \eqref{eq:masterformulaintro}  to any toric Calabi-Yau threefold, in any patch of the closed
string moduli space, also away from, and possibly in absence of, orbifold points. Secondarily, we exhibit a
direct connection of the closed, $T$--equivariant theory with descendants in
the r.h.s. of \eqref{eq:masterformulaintro} with
the theory of classical Hamiltonian integrable systems in 1+1
dimensions. \\

The resulting formalism, which is perturbative in $g_s$ but which holds true
globally in $\a'$, can be regarded as 
%the proposal of 
an A-model mirror of the remodeled-B-model
of \cite{Bouchard:2007ys}, and is interesting for a number of reasons. First of all,
from a conceptual point of view, it gives a purely A-model formulation of the
problem of computing $F^{X,L}_{g,h}$, which coincides with their rigorous (albeit
purely calculational) 
mathematical definition from localization. Second,  it is valid
for general toric Calabi-Yau threefolds, and in all chambers of the extended K\"ahler moduli
space of $X$, including orbifold points. Thirdly, while
not being as computationally straightforward as the BKMP-formalism
\cite{Bouchard:2007ys}, it is still surprisingly effective for computing the type of
amplitudes considered in \cite{Bouchard:2007ys}, especially considering the fact
that its starting point is completely rooted on the A-side, and in some
examples it goes beyond the methods known to date. Fourthly, it exhibits a novel, clear
connection to underlying integrable structures of the topological string; the
resulting picture is quite
different from (and in a way simpler than) the one arising from the dual Kodaira-Spencer theory on the
B-model side \cite{Aganagic:2003qj}, and it provides moreover a clear identification of a key object in
the relation of topological strings to matrix models, namely, the brane insertion
operator $\LL(w,f)$. Fifthly, it embeds in a systematic fashion the change of the canonical choice
of flat open string moduli when moving from one chamber to another of the
K\"ahler moduli space. Finally, it can be used to make contact
with the results of local mirror symmetry, and most notably to recover the
mirror Calabi-Yau geometry based on spectral curves from a resummation of
A-model instantons. \\

The paper is organized as follows. We first review in Sec.~\ref{sec:review} the necessary background on
the open and closed topological A-model on toric Calabi-Yau threefolds; in
view of the role of the master formula \eqref{eq:masterformulaintro}, we
discuss in some detail the closed equivariant side and the ``recoupling'' to
topological gravity induced by gauging the $T$-action. We then describe our
formalism in Sec.~\ref{sec:formalism}: we first review the derivation of
\eqref{eq:masterformulaintro} for orbifolds and formulate its extension to
general toric Calabi-Yau threefolds. We then describe the relation of the
closed equivariant model with integrable hierarchies, and discuss its general concrete
implementation at low genera. A crucial role here is played by Dubrovin's
theory \cite{Dubrovin:1992dz, Dubrovin:1994hc} of dispersionless hierarchies arising from associativity equations,
and their dispersive deformation \cite{dubrovin-2001} via the group of rational Miura
transformations. Section~\ref{sec:examples} is devoted to our three main
examples: the framed vertex, for which we find a relation to
a disguised form of the KdV hierarchy, the resolved conifold, where the
relevant integrable system is the Toeplitz reduction of the 2D-Toda hierarchy
\cite{Brini:2010ap, ioguidopaolo}, 
and local $\bbP^2$, where we explicitly test that our formalism correctly
computes topological amplitudes possessing a non-trivial
quasi-modular dependence on the closed string moduli; we also briefly report on the
case of a particular $\bbZ_7$ orbifold of $\bbC^3$, for which computations of
open string amplitudes at the orbifold point would be awkward (if not
unfeasible) with other methods. In
Sec.~\ref{sec:instantons} we make contact with toric mirror symmetry and
derive in each of our examples 
the Hori-Vafa spectral curves by summing over open string instantons at $g=0$,
$h=1$. We conclude in Sec.~\ref{sec:conclusions} with some remarks on new possible
developments. Some background material on $I$-functions of toric orbifolds
%, as
%well as the details of the local $\bbP^2$ orbifold phase transition in our
%language, 
are included in the appendix.

%\vskip 1truecm
% \noindent {\large{\bf Acknowledgements}}
%\\
\section*{Acknowledgements}
\hyphenation{Tan-zi-ni}

\noindent I am particularly indebted to R.~Cavalieri and T.~Coates for many insightful
discussions and patient explanations. I would also like to thank M.~Aganagic,
G.~Bonelli, D.~E.~Diaconescu,
B.~Dubrovin, 
B.~Eynard, K.~Hori, A-K.~Kashani-Poor, A.~Klemm, M.~Mari\~no, A.~Oblomkov,
S.~Pasquetti, A.~Tanzini, Y.~Zhang
for discussions and/or comments on the manuscript, as well as the Geometry group of Imperial College, the LPTENS,
the Department of Mathematics of Colorado State University and the Theory
groups at Berkeley, Caltech and Rutgers for their hospitality while this work
was being finished. This work was supported by a Postdoctoral Fellowship of the Swiss
National Science Foundation (FNS).

%\bit
%\item novel, pure A-model
%\item global, general, works!
%\item neat connection to integrable hierarchies
%\item hidden relation to matrix models via LIO
%\item systematizes choice of flat open string moduli
%\item spectral curves from resummation of disc instantons
%\eit

\section{The open and closed A-model on toric Calabi-Yau threefolds}
\label{sec:review}
\subsection{The open string side}
%{\bf due parole su A-model aperto in generale}

\subsubsection{Geometry}

We will be interested in the topological A-model on a
toric Calabi-Yau threefold (TCY3) $X$ with a background Lagrangian toric
brane $L\hookrightarrow X$; we briefly review in this section the geometric
setup. There is no new material here; further details may be found in
\cite{Aganagic:2000gs, Aganagic:2001nx, Bouchard:2007ys, kl:open, MR2003030}. \\

\begin{figure}[!h]
\begin{minipage}[t]{0.46\linewidth}
\centering
\vspace{0pt} 
\includegraphics[scale=0.6]{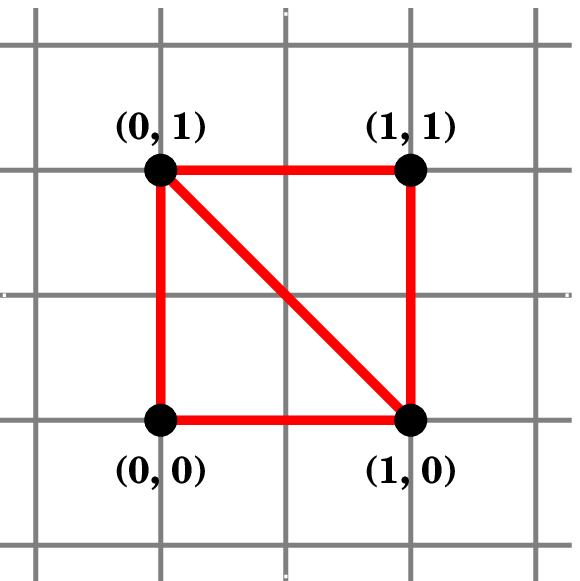}
\vspace{0pt} %\hspace{-10em}
\caption{\small The toric diagram $\Sigma_X$ of $\cO_{\bbP^1}(-1)\oplus\cO_{\bbP^1}(-1)$.}
\label{fig:sigmax}
\vspace{1.5cm}
\end{minipage} 
\vspace{1cm}
\begin{minipage}[t]{0.08\linewidth}
\end{minipage}
\begin{minipage}[t]{0.46\linewidth}
\centering
\vspace{.5cm} 
\includegraphics[scale=0.8]{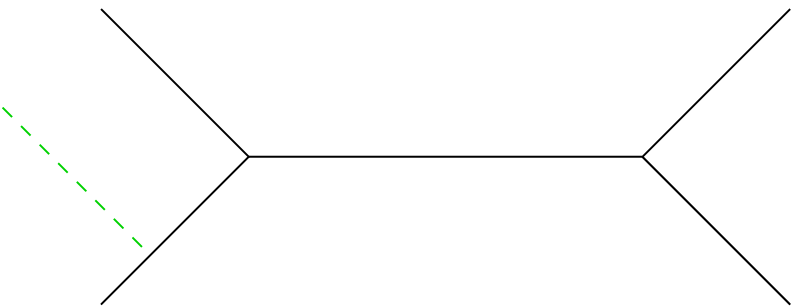}
\vspace{.6cm} 
%\hspace{-10em}
\caption{\small The web diagram of $X=\cO_{\bbP^1}(-1)\oplus\cO_{\bbP^1}(-1)$,
with a brane on an external leg.}
\label{fig:pqwebx}
\end{minipage} 
\end{figure}

By definition, a smooth TCY3 $X$ is a K\"ahler manifold with vanishing canonical class and admitting
a complex rank three group of holomorphic isometries, whose (algebraic) maximal
torus we denote by $S\simeq (\bbC^*)^3$. This last fact allows to describe $X$
in a purely diagrammatic way in terms of a three-dimensional integer
sublattice of $\bbZ^3$ - the {\it fan} $\cF_X$
of $X$ - which specifies the way the $S$-orbits close (see \cite{MR1677117,
  MR2003030} for details). In the Calabi-Yau case which is of our interest, this information can be
compactly encoded in a triangulated polytope $\Sigma_X\subset \bbZ^2$ - the {\it
  toric diagram} of $X$, as in Figure \ref{fig:sigmax}.
Alternatively \cite{Aganagic:2003db}, and equivalently, $X$ can be realized as
a degenerate $\bbT^2\times \bbR_+$ fibration over $\bbR^3$; the geometry is
completely specified by a finite number of integer data which specify the
degeneration loci of the torus fibers. The degeneration locus, which is the
image of the moment map by which $X$ is realized, has an easy
interpretation in terms of the Newton polytope $\Sigma_X$: it is the dual
polytope of $\Sigma_X$, and goes under the name of the {\it web diagram} of $X$. See Fig. \ref{fig:pqwebx}. \\

We now want to turn on an open string sector. In the toric setting there is a
distinguished set of Special Lagrangian submanifolds $L \simeq \bbR^2 \times
S^1 \hookrightarrow X$; these are often referred
to as ``toric branes'' and were constructed in \cite{Aganagic:2000gs}
generalizing \cite{MR666108, Ooguri:1999bv}. In terms of the  web diagram, $L$ is
constructed as the co-normal bundle over a straight semi-infinite line in the toric
polytope, intersecting one of its edges. To visualize it,  let $v$ be a vertex
of the toric web and $l_i$, $i=1,\dots,3$ be the three outcoming
edges\footnote{We keep supposing that $X$ is smooth, hence all vertices of the
  toric web are trivalent.}, and let $L$ intersect the edge $l_3$. In local
co-ordinates $x_1$, $x_2$, $x_3$ dual to the faces $l_{23}$, $l_{13}$ and
$l_{12}$ of the web, the Lagrangian $L$ is the fixed locus of the
anti-holomorphic involution $\sigma:X\to X$ sending
\beq
\sigma : (x_1,x_2,x_3) \to \l(\bar x_3 \bar x_2, \bar x_3 \bar x_1,1 / \bar
x_3\r)
\label{eq:involution}
\eeq 
The $S^1$ inside $L$ is the equator $|x_3|=1$ of the $x_3$ direction, whereas the
$\bbR^2$ factor comes from the two real fibers in the transverse direction as in \eqref{eq:involution}.\\

\subsubsection{A-model open string instantons}
The type A open topological string on $(X,L)$ deals with the (virtual) count of the
number of A-model instantons  from a genus $g$, $h$-holed worldsheet
$\Gamma_{g,h}$ to $X$, with Dirichlet boundary conditions given by $L$. From a
geometric point of view, the natural object to parameterize such
instantons would be ideally given by a suitable compactification of a moduli
space $\cM_{g,h}(X,L,\beta,\{ d_i \}_{i=1}^h)$  of maps
$\phi:\Gamma_{g,h}\to X$, such that $\phi$ is holomorphic in the bulk and is
continuous along the boundary $\de
\Gamma_{g,h} \subset L$, in a fixed topological sector specified by
$\phi_*[\Gamma_{g,h}]=\beta\in H_2(X,L, \bbZ)$ and $\phi_*[D_i]=d_i\in
H_1(L)\simeq \bbZ$. The genus $g$, $h$-holes contribution to the open topological string
partition function $Z^{X,L}$ would then be given as
\beq
(\log Z^{X,L})_{g,h}({\bf t}, {\bf w})= F^{X,L}_{g,h}({\bf t}, {\bf w}) =
\sum_{\beta \in H_2(X,L,\bbZ)}\sum_{d_1, \dots, d_h\in \bbZ}
N^{X,L}_{g,h,\beta,d_1, \dots, d_h} \re^{t\cdot \beta} \prod_{i=1}^h w_i^{d_i} 
\eeq
where ``open Gromov-Witten invariants'' $N^{X,L}_{g,h,\beta,d_1, \dots, d_h}$
would be defined as
\beq
N^{X,L}_{g,h,\beta,d_1, \dots, d_h} \stackrel{!}{=} \int_{[\overline{\cM}_{g,h}(X,L,\beta,\{
    d_i \}_{i=1}^h)]^{\rm vir}}1
\label{eq:opengw}
\eeq
The definition, and {\it a fortiori} the calculation of
$N^{X,L}_{g,h,\beta,d_1, \dots, d_h}$ in \eqref{eq:opengw} hinges on finding a
suitable Kontsevich-like compactification of $\overline{\cM}_{g,h}(X,L,\beta,\{
    d_i \}_{i=1}^h)$ and on the construction of a top-dimensional homology
    cycle - the {\it virtual fundamental class \cite{MR1437495}} - with the properties expected from deformation theory. 
While the conceptual discussion for the open topological A-model would parallel
the ordinary closed string case, there are however several important technical points where the
open case departs from the one without $D$-branes. \\

To start with, the real
condition imposed by the Lagrangian forces the open string moduli spaces to be
essentially non-(complex)-algebraic, and makes it more difficult to find a
viable mathematical compactification of the moduli space as compared to the case of closed
strings, let alone the construction of a virtual fundamental
class. As emphasized in the foundational work \cite{Solomon:2006dx}, the construction of open string
moduli spaces for $(X,L)$ naturally leads to problems coming from the
non-orientedness of $\overline{\cM}_{g,h}(X,L,\beta,\{
    d_i \}_{i=1}^h)$, and to the fact that it has a non-trivial boundary in
    real co-dimension one, thus making open string insertions in principle ill-defined at
    the level of co-homology\footnote{The reader is referred to
      \cite{Solomon:2006dx}, where these problems are addressed in the case of
      branes described as the fixed locus of an anti-symplectic
      involution. For the toric case with Aganagic-Vafa branes, see also
\cite{Li:2001sg} for an algebraic definition in terms of moduli spaces of
relative stable morphisms.}. \\
One possible way to circumvent this problem and define operatively the
invariants is to use localization, an approach put forward in the work of Katz and Liu
\cite{kl:open}. Recall that in the closed string case we do have (at least in positive degree) a well-defined construction of a virtual
fundamental cycle $[\overline{\cM}_{g}(X,\beta)]^{\rm vir}$ for the moduli
space of stable maps to $X$, for example in terms of relative stable maps to a
projective compactification of $X$. 
%or, in the case of concave bundles over a
%projective toric curve/surface, of twisted Gromov-Witten invariants of its
%null section. 
A torus action $T \times X \to X$ on $X$ pulls back to an action
on $\overline{\cM}_{g}(X,\beta)$; the fundamental cycle
$[\overline{\cM}_{g}(X,\beta)]^{\rm vir}$ induces a $T$-equivariant virtual cycle
$[\overline{\cM}^T_{g}(X,\beta)]^{\rm vir}$ on the $T$-fixed locus of
  $\overline{\cM}_{g}(X,\beta)$. Closed string Gromov-Witten invariants are then computed as
\beq
N^{X}_{g,\beta}:= \int_{[\overline{\cM}^T_{g}(X,\beta)]^{\rm vir}}1 = \sum_i \int_{[\gamma_{i,g,\beta}]^{\rm vir}}\frac{1}{\re_T(N^{\rm vir}_{\gamma_{i,g,\beta}})}.
\label{eq:equivint}
\eeq
where we denoted by $\gamma_{i,g,\beta}$ the fixed components of the $T$-action. In presence of a torus action $T\simeq\bbC^*$ compatible with the
anti-holomorphic involution defining $L$, an
extension of this line of reasoning to the open string setting was given in
\cite{kl:open}. The authors propose a natural tangent/obstruction theory for
the moduli space of open stable maps; the relevant exact sequence reads, in
terms of fibers at a smooth point $(\Gamma,f)$
\begin{align}\label{ot}
0\to H^0(\Gamma,\partial \Gamma,T_\Gamma, T_{\partial\Gamma}) \to
H^0(X,L,f^\ast T_X, (f_{|\partial\Gamma})^\ast T_L) \to \mathcal{T}^1 \to \nonumber \\
 H^1(\Gamma,\partial \Gamma,T_\Gamma, T_{\partial\Gamma}) \to
H^1(X,L,f^\ast T_X, (f_{|\partial\Gamma})^\ast T_L) \to \mathcal{T}^2 \to 0
\end{align}
The resulting moduli space, when $X$ is a CY3 and $L$ the fixed locus of an
anti-holomorphic involution, has expected dimension zero. If we now {\it
  assume}
that 
\ben 
\item there is a well-defined $T\simeq\bbC^*$ action on the moduli space, so
  that localization theorems apply;
\item we can identify the $T$-fixed loci $\gamma_{i,g,\beta,\{ d_i
  \}_{i=1}^h}$, and have a natural proposal for the 
%existence and a
%computational definition of the 
localization of the fundamental cycle
$1^{\rm vir}_{\gamma_{i,g,\beta,\{ d_i \}_{i=1}^h}}$,
\een
then the open Gromov-Witten invariants \eqref{eq:opengw} can be {\it defined} by
localization
\beq
N^{X,L}_{g,h,\beta,d_1, \dots, d_h}  := \sum_i \int_{[\gamma_{i,g,\beta,\{ d_i \}_{i=1}^h}]^{\rm vir}}\frac{1}{\re_T(N^{\rm vir}_{\gamma_{i,g,\beta,\{ d_i \}_{i=1}^h}})}.
\label{eq:opengwloc}
\eeq
The point which is harder to prove rigorously is the first. If we assume this,
though, \eqref{eq:opengwloc} yields an operative definition of the open
A-model on $(X,L)$. In particular, in the case of toric backgrounds with
Aganagic-Vafa branes and in presence of a Calabi-Yau action $T$ compatible with $L$, it is easy to determine the topological data that define
the localization of the virtual cycle to the $T$-fixed loci. As the choice of
the torus  $T$ is non-unique, but rather
depends on an integer ambiguity $f\in \bbZ$, the resulting open string
invariants depend on an additional $\bbZ$-valued parameter
\beq
N^{X,L}_{g,h,\beta,d_1, \dots, d_h} = N^{X,L}_{g,h,\beta,d_1, \dots, d_h}(f)
\eeq
This fact is entirely expected from string duality, as it corresponds to the large $N$ dual incarnation of the
framing ambiguity of Wilson loops of knots and links in Chern-Simons theory
\cite{Witten:1988hf, Ooguri:1999bv}. \\
%IR ambiguity check!!!!

In Sec.~\ref{sec:orbifolds} we will review the structure of the Atiyah-Bott
computations behind \eqref{eq:opengwloc}. To conclude this section, let us
just mention on the case where $X$ is not smooth. The picture can in fact be
generalized to include singular toric Calabi-Yau threefolds
\cite{Bouchard:2007ys, Bouchard:2008gu, Brini:2010sw}. In particular, for $G$
finite abelian, let $X=[\bbC^3/G]$ be toric Calabi-Yau  orbifold of
 flat space; we choose the
fibers $x_i$, $i=1,2,3$ to carry irreducible representations of the
$G$-action. As the anti-holomorphic involution \eqref{eq:involution} is
compatible with a Calabi-Yau $G$-action, it
descends to the quotient defining a Lagrangian $L\subset  \bbC^3/G$; in
presence further of a compatible $T$-action, localization can be applied to
define/compute open orbifold Gromov-Witten invariants of the pair $(X, L)$
in this more general case.

\subsection{The equivariant closed string side}
\label{sec:closed}
As we will see, closed string localization formulae will play a crucial role
in what follows; we will then be interested in the problem of computing 
$T$-equivariant Gromov-Witten invariants of $X$. Since turning on a torus action leads to a number of new
interesting phenomena with respect to the ordinary non-equivariant case, we
briefly review them here. \\
\subsubsection{The $T$-equivariant A-model}
At a worldsheet level, the equivariant integration on the closed string moduli
spaces $\overline \cM_{g}(X,\beta)$ is realized as follows. Since $X$ is a
toric threefold, we have $\mathrm{rank}_{\bbC} \mathrm{Iso}(X)= 3$. Whenever the target space
possesses a flavor symmetry in the form of a holomorphic isometry, this can be used to generate a $\cN=(2,2)$ potential deformation
of the original worldsheet theory  \cite{MR719813, Labastida:1991yq, Labastida:1996tz}. Let $\phi^i$ be local
charts on $X$, $V \in \mathfrak{iso}(X)$ and write $V=V^i \de_{\phi_i}$ in
components; we will write $T\subset \mathrm{Diff}(X)$ for the the abelian flow
generated by $V$.  
%The deformation leading to \eqref{eq:equivint} is obtained by
%gauging the flavor symmetry  generated by $X$,  going to the weak gauge
%  coupling limit and freezing the new vectors
%\footnote{We sometimes refer to this model as the ``gauged''
 %   A-model, but we emphasize once again that we are really not just gauging a flavor
 %   symmetry here. Not freezing the gauge degrees of freedom 
%    \cite{Hanany:1997vm, Hori:2000kt} would result in a different 2D-vortex-counting
%  problem \cite{MR1777853, Baptista:2007ap};  a moduli
%  space picture for the decoupling limit leading to equivariant Gromov-Witten
%  theory was recently given in
%  \cite{Gonzalez:2009fu}.} 
%at $A_\mu=0$, $\psi=0$,
%  $\phi=\lambda\in \bbC$. 
In terms of worldsheet fields in the untwisted
  theory the deformation reads
\beq
\delta_T L = - g_{i \bar j} |\lambda|^2 V^i \bar V^{\bar
  j}-\frac{i}{2}\l(g_{i\bar i}\de_j V^i-g_{j\bar j}\de_i V^j\r)\l(\lambda
\psi^{\bar i}_- \psi_+^j + \mathrm{h.c.}\r).  
\label{eq:deform}
\eeq
where the complex masses $\lambda$ are the equivariant parameters of the $T$-action.\\ 
This deformation has a series of important consequences. The theory has a modified $\cN=(2,2)$ supersymmetry with a non-vanishing
central extension given by $\mathrm{Lie}_V$; as a consequence, in the A-topologically twisted
theory the BRST differential $Q$ is deformed to the equivariant de Rham
differential $d-\sqrt{2}i\lambda i_V$, hence squaring to $Q^2=2i\lambda
\mathrm{Lie}_V$. \\

A-model chiral operators $\cO_\a$ from the $\sigma$-model sector of the theory are now in one to one orrespondence
to {\it invariant} forms $\cO_\a\leftrightarrow \phi_{\a} \in H^\bullet_T(X)$. At the level of the corresponding moduli space of classical trajectories, chiral $n$-point functions are computed as
\beq
\bra \cO_{\a_1} \dots  \cO_{\a_n} \ket^{X_T}_{g,n,\beta} = \sum_i \int_{[\gamma_{i,g,\beta,n}]^{\rm vir}}\frac{\prod_{i=1}^n\rm{ev}_i^*\phi_{\a_i}}{\re_T\Big(N^{\rm vir}_{\gamma_{i,g,\beta,n}}\Big)}.
\label{eq:corr1}
\eeq
where the moduli spaces parametrize stable maps from $n$-pointed curves,
with markings corresponding to chiral insertions; we wrote
$\mathrm{ev}_i:\overline{\cM}_{g,n}(X,\beta) \to X$ for the evaluation
morphism at the $i^{\rm th}$ marked point
\beq
\mathrm{ev}_i([\Gamma, f,p_1, \dots, p_n])= f(p_i). 
\eeq
As for the ordinary A-model, the resulting chiral ring is an $\a'$-deformation of the  $T$-equivariant de
Rham co-homology $H^\bullet_T(X)$, called the {\it big quantum co-homology
ring}. When we want to emphasize the stringy deformation of the ring structure,
will write $QH_T^\bullet(X)$ to denote the $T$-equivariant chiral ring. \\ 

As a further comment, notice that the
extra term \eqref{eq:deform} results in a deformation of the
worldsheet theory away from conformality, giving new (twisted) mass terms for the
fermions. Moreover, the new fermion mass terms have charge 2 under the A-model
ghost number charge, whose R-symmetry is thus broken explicitly by
\eqref{eq:deform}. From a spacetime point of view, the resulting topological string yields refined invariants, counting
the number of wrapped $M2$ branes of definite charge under the $T$-action (see
\cite{Aganagic:2002qg} for a discussion, as well as a large $N$ dual
description in Chern-Simons theory).  \\

\subsubsection{Recoupling to the observables of topological gravity}
Gauging the torus action
results in an important  new extra feature \cite{Hanany:1997vm, Hori:2000kt}: in
the full topological string, we have an infinite tower of non-trivial BRST-closed descendants
$\cO_{\a,p}$ for each  
$\cO_\a$, involving the observables of the gravitational sector of the theory
(see e.g. \cite{MR1068086, Dijkgraaf:1990qw}). They have the form
\beq
\cO_{\a,p}=\sigma^p \phi_\a
\eeq
where $\sigma$ is the superfield obtained by topological descent equations on
a bosonic operator $\sigma^{(0)}$ given in conformal gauge by
\beq
\sigma^{(0)}  = \frac{1}{2}\l(\de \gamma+\gamma\de\phi-c\de \psi - \mathrm{h.c.}\r)
\eeq
in terms of the ghost $\beta$, $\gamma$, the 2D gravitino $\psi$, and the
Liouville field $\phi$. Their moduli space realization is given in terms of
powers of tautological classes \cite{MR975222}: this leads to a $\bbZ^n$-family of $n$-point chiral {\it gravitational} correlators
\beq
\bra \cO_{\a_1,p_1} \dots  \cO_{\a_n, p_n} \ket^{X_T}_{g,n,\beta} =
\sum_i \int_{[\gamma_{i,g,n,\beta}]^{\rm
    vir}}\frac{\prod_{i=1}^n\rm{ev}_i^*\phi_{i} \psi^{p_i}}{\re_T\Big(N^{\rm vir}_{\gamma_{i,g,n,\beta}}\Big)}.
\label{eq:corr1grav}
\eeq
obtained by capping the pull-backs at the $i^{\rm th}$ marked point of
$\phi_{\a}$ with powers of the $i^{\rm th}$ tautological class
$\psi_i=c_1(\bbL_i)$, where $\bbL_i$ is the line bundle on
$\overline{\cM}_{g,n}(X,\beta)$ whose fiber over a smooth moduli point
$(f,\Gamma, p_1, \dots, p_n)$ is the cotangent line $T^*_{p_i}\Gamma$. \\

In the ordinary
Calabi-Yau case, this infinite set of gravitational operators is largely
decoupled, and at any rate it does not contain any
new information with respect to the partition function: $U(1)_R$ charge
conservation forces this type of insertions to be mostly zero, or to be
trivially proportional to the free energy. In the $T$-equivariant case, instead, the Calabi-Yau
selection rules that ``decoupled'' topological gravity are violated by
terms proportional to the mass terms in \eqref{eq:deform}, namely, the equivariant parameter $\lambda =
c_1(\cO_{\bbC\bbP^\infty}(1))   \in H_T(\{\mathrm{pt}\})$ of $T\times X \to
X$. This means that the gravitational correlator \eqref{eq:corr1grav} is now generically non-vanishing and carries extra
information (proportional to the $T$-generated mass terms) with respect to the
partition function. As a consequence, the equivariant topological A-model on a
toric Calabi-Yau threefold closely resembles the topological string in the
asymptotically free case, such as the A-model on Fano target manifolds. \\

As for the ordinary non-equivariant case, it is convenient to pack together equivariant
Gromov-Witten invariants inside generating functions. To this aim, we first of
all introduce chemical
potentials $t_{\a,p}$ dual to insertions of $\cO_{\a,p}$ and we write the
genus $g$, full-descendant equivariant A-model free energy as
\beq
\cF^{X,T}_g(t_{\a,p})=\sum_{\beta\in
  H_2(X,\bbZ)}\sum_{n=0}^\infty\sum_{\stackrel{\a_1, \dots, \a_n}{p_1,
    \dots, p_n}} \frac{\prod_{i=1}^nt_{\a_i,p_i}}{n!}\bra  \cO_{\a_1,p_1} \dots  \cO_{\a_n, p_n}  \ket^{X_T}_{g,n,\beta}
\label{eq:fulldesc}
\eeq
where $\a=1,\dots,\chi(X)$ and $p_i\in \bbZ^+$ for all $i$; we use the fact
that $\dim_{\bbC(\lambda)} H_T(X)=\chi(X)$ in the toric case, due to the
vanishing of the co-homologies in odd degree. The
non-equivariant free energies $F^{X}_g(t_{\a})$ can be recovered in two
ways\footnote{To be rigorous, both statements hold strictly speaking only when we discard unstable
contributions to the free energy and regularize the degree zero terms at
tree-level, as the latter are necessarily singular non-equivariantly due to the
non-compactness of $X$.}: either by taking $\lambda\to 0$, or, when the
$T$-action preserves the Calabi-Yau condition, by restricting the insertions
to {\it small quantum co-homology}, namely, to degree 2 primary operators
\beq
F^{X}_g(t_{\a})=\cF^{X,T}_g(t_{\a,p})\Big|_{t_{\a,p}=t_{\a}\delta_{p,0} \delta_{\mathrm{deg}\a,2}}
\eeq
An important role in our formalism will be played by a restriction of
\eqref{eq:fulldesc} to its $g=0$, $n=1$ subsector, in the form of Givental's
$J$-function. This is the $H^\bullet_T(X)$-valued power series
\beq
J^{X,T} \l(t^1, \dots, t^{\chi(X)};  z\r) := z + t_\a\phi^\a +
\sum_{n=0}^\infty\sum_{\beta\in H_2(X,\bbZ)}\bra\mathbf{t}, \dots, \mathbf{t},
\frac{\phi_\a}{z-\psi}\ket^{X_T}_{0,n+1,\beta} \phi^\a
\label{eq:Jfun}
\eeq
where 
\beq
\mathbf{t}:=\sum_{\beta=1}^{\chi_X}t_\beta \phi^\beta.
\label{eq:sumt}
\eeq
Restricting the sum in \eqref{eq:sumt} to degree $\leq 2$ classes, we obtain the {\it
  small} J-function of $X$.
\section{Our formalism}
\label{sec:formalism}
Our formalism builds directly on the localization approach to
the open A-model on toric backgrounds \cite{kl:open, Brini:2010sw}. In the
following, we first of all recall the relation between open string amplitudes
and closed descendant invariants in the form \eqref{eq:masterformulaintro},
and give new arguments for its general validity, also in singular phases. We
then review and build on the connection between the closed equivariant theory
and $\tau$-functions of classical integrable hierarchies.
\subsection{Localization and an open/closed string duality}
In this section we outline the derivation of the master formula
\eqref{eq:masterformulaintro}. We first concentrate on the case when $X$ is a
toric orbifold of $\bbC^3$, and then generalize \eqref{eq:masterformulaintro}
to an arbitrary open string toric background $(X,L)$. 
\subsubsection{Orbifolds}
\begin{figure}[t]
\centering
\includegraphics[scale=0.6]{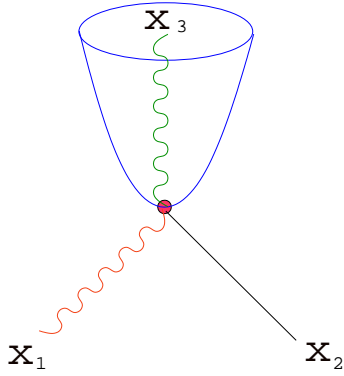}
\vspace{.6cm} 
%\hspace{-10em}
\caption{\small  The orbifold vertex  $X=[\bbC^3/G]$, depicted with two gerby
  lines along $x_1$ and $x_3$, and a Lagrangian
  intersecting the equator of the $x_3$ fiber.}
\label{fig:orbvertex}
\end{figure}

For future utility, before moving to the analysis of open string instantons for
this brane setup let us recall the structure of the closed string sector. To start with, suppose $X$ is a Calabi-Yau abelian orbifold
of $\bbC^3$, and let $L\hookrightarrow X$ be the Lagrangian defined by
\eqref{eq:involution} (see Fig. \ref{fig:orbvertex}). 
Consider the following family of $T\simeq \bbC^*$ actions on $X$
\beq
\bary{cccc}
T \times  X & \to & X\\
(\mu, x_1, x_2,x_3) & \to & (\mu^{f} x_1, \mu^{-f-r_{\mathrm{eff}}}
  x_2, \mu^{r_{\mathrm{eff}}} x_3)
\eary
\label{eq:torusaction}
\eeq
In \eqref{eq:torusaction}, $r_{\mathrm{eff}}=\mathrm{l.c.m.}(p^{\mathrm{eff}}_1, \dots,
p^{\mathrm{eff}}_r)$, where we decomposed $G=\times_{i=1}^r \bbZ_i$ and we
denoted by $p^{\mathrm{eff}}_i=\mathrm{ord}(\bbZ^{\mathrm{eff}}_i)$ the order of
the maximal subgroup of the $i^{\rm th}$ $\bbZ_i$ factor acting effectively
along the $x_3$ fiber. The framing parameter $f$ can in principle be rational
when $G\neq e$, with a denominator that divides $r_{\rm eff}$. \\

The $T$-equivariant chiral ring in this case coincides classically
\cite{Zaslow:1992rp, Chen:2000cy} with the
$T$-equivariant orbifold co-homology $H^\bullet_{\mathrm{orb},T}(X)$
\beq
H^\bullet_{\mathrm{orb},T}(X)=\bigoplus_{g\in G} H^\bullet_{T}(X_g)
\label{eq:orbcoh}
\eeq 
where $X_g=\{(x,g) | x \in \bbC^3\}$ denotes the $g$-twisted sector of $X$
for $g\in G$. We write $\cI X:=\bigsqcup_{g \in
  G} X_g$ for their disjoint union - the {\it inertia stack} of $X$ -
%  $I$ for the $\bbZ_2$-automorphism of $\cI X$ sending $(x,g) \to (x,
%  g^{-1})$, 
and $\mathbf{1}_g$ for a
$G$-twisted\footnote{We
  hasten to warn the reader that by ``twisted'' we do {\it not} mean ``twisted chiral'': all
  the discussion is strictly holomorphic here. In this section, 
  ``twisted'' refers to the fact that these insertions come from the twisted sectors of the orbifold
  topological string.} class in $H^\bullet_{\mathrm{orb},T}(X)$.  The topological two point
function is given by the $T$-equivariant orbifold Poincar\'e pairing \cite{Chen:2000cy}
\beq
\eta(\mathbf{1}_g, \mathbf{1}_{g'})=\int_{\cI X^T} \frac{\mathbf{1}_g \cup
  \mathbf{1}_{(g')^{-1}}}{e(N_{\cI X^T/ \cI X})}
\eeq
The A-model ghost number charge in a twisted sector has a contribution coming
from the vacuum fermionic shift that affects the orbifold topological string
in presence of $g$-twisted boundary conditions \cite{Zaslow:1992rp}. Without
loss of generality, suppose $x_i$, $i=1,2,3$ carry one-dimensional irreps of
$G$ and denote with $\a^{r}_{g,i}$ the character of $g=(g_1, \dots, g_r) \in
\times_{i=1}^r \bbZ_i$ on the line parameterized
by $x_i$. Then the fermionic shift is equal to twice the {\it age} of
$\mathbf{1}_g$  \cite{Zaslow:1992rp, Chen:2000cy}
\beq
\mathrm{age}(\mathbf{1}_g)=\sum_{i=1}^3 \sum_{j=1}^r\frac{\a^{j}_{g,i}}{p^{\rm
  eff}_{j}} \in \bbZ
\eeq
and we have for the orbifold (or Chen-Ruan) degree
\beq
\mathrm{deg}(\mathbf{1}_g)=2\mathrm{age}(\mathbf{1}_g)
\eeq

\subsubsection{The master formula for open string invariants: the orbifold vertex}
\label{sec:orbifolds}

\begin{figure}[t]
\centering
\includegraphics[scale=0.6]{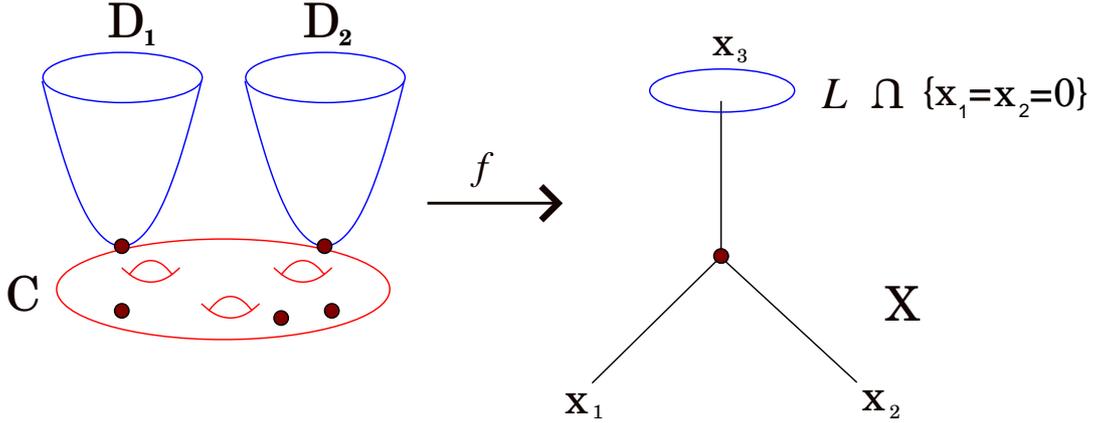}
\vspace{.6cm} 
%\hspace{-10em}
\caption{\small  A pictorial view of a torus-fixed stable map from a genus
  $g=3$, $h=2$ (orbi)-curve  to  $X=[\bbC^3/G]$ (from \cite{Brini:2010sw}).}
\label{fig:fixlocus}
\end{figure}

Open string instantons of $(X, L)$ have the following structure.
The $T$-fixed points inside $\overline{\cM}_{g,h, n}(X,L,\beta,\{ d_i \}_{i=1}^h)$ consist
of a compact genus $g$ curve, carrying $n$ twisted marked points, with a
collection of $h$ (orbi-)discs attached, as depicted in Figure
\ref{fig:fixlocus}. 
The compact curve contracts to the vertex of the toric diagram, and the discs
are mapped (rigidly) to the lower hemisphere $|x_3|<1$ of the compactified
$x_3$ fiber, with their boundary wrapping around the equator. \\
The restriction of the Katz-Liu obstruction theory \eqref{ot} to a
$T$-fixed locus consists of essentially three pieces; the reader is referred
to \cite{kl:open, Brini:2010sw} for more details. Suppose we only have
primary (matter) insertions $\phi_{\a_i} \in H^\bullet_{\rm orb}(X)$,
$i=1,\dots, n$, of Chen-Ruan degree 2 at the $n$ marked
points. The contracting compact curve and the $n$ marks yield a factor
\beq
\prod_{i=1}^3
\Lambda_i(\mu_i)\prod_{j=1}^n \mathrm{ev}_j^* \phi_{\a_j} \in
H^{\mathrm{top}}(\overline{\cM}_{g,n}(B G,0), \bbQ)
%=\prod_{i=1}^n\phi_{\a_i}\ket_{g,n,\beta=0}^{X_T}
\label{eq:compact}
\eeq
corresponding to the dual of an $n$-pointed insertion that contributes to the
closed equivariant genus $g$ free energy; in \eqref{eq:compact},
$\Lambda_i(\mu_i)$ is the $T$-equivariant Euler class of the dual of an
appropriate sub-bundle of the Hodge bundle \cite{Brini:2010sw}, linearized with the weights of the
torus action $\mu_1=-f-r_{\mathrm{eff}}$, $\mu_2=f$, $\mu_3=r_{\mathrm{eff}}$
as in \eqref{eq:torusaction}. The $i^{\rm th}$ node contribution brings about a
gravitational contribution of the type $\frac{1}{\frac{\lambda}{d_i}- \psi }$
for each node $i$, as well as a universal constant normalization term \cite{Brini:2010sw};
finally, each disc contributes a factor $D^{X,L}_{\a}(d_i,f)$, which is completely
determined by the brane setup and the choice of framing. Altogether, and
taking into account \cite{Brini:2010sw} the compatibility condition
between degree of the map and twisting at the nodes, we obtain that a genus
$g$, $h$-holed amplitude on $(X, L)$ is given by\footnote{In some
  conventions \cite{Graber:2001dw},
$F^{X,L}_{g,h}$ is an element in $H_T^{2h}(\mathrm{pt})$; what we call 
  $F^{X,L}_{g,h}$ here is the coefficient of proportionality of $\lambda^{h}$ there.}
\beq
F^{X,L}_{g,h}(t_1, \dots, t_n; w_1, \dots w_h;f) = \l(\prod_{i=1}^h \LL^{X,L}(w_i,f)\r)
\cF^X_{g}(t_{\a,p}, f)\Bigg|_{t_{\a,p}=\delta_{{\rm deg\phi_\a},2 }
  \delta_{p,0}t_\a}.
\label{eq:masterformulaorbifolds}
\eeq
where the {\it brane insertion operator} $\LL^{X,L}(w,f)$ has the form
\beq
\LL^{X,L}(w,f)=\sum_{n=0}^\infty\sum_{d=1}^{\infty}\sum_{\a\in
  H_{\rm orb}^\bullet(X)} w^d d^{n+1}
D^{X,L}_\a(d,f)\frac{\de}{\de t_{\a,n}}
\label{eq:lio}
\eeq
and $\cF^X_{g}(\tau_{\a,p}, f)$ is the full-descendant genus $g$ free energy
\eqref{eq:fulldesc}. This is what we called the ``master formula'' for open
string amplitudes in \eqref{eq:masterformulaintro}.\\

The disc contribution $D^{X,L}_\a\left(d, f\right)$ is what specifies the form
of the brane insertion operator $\LL^{X,L}(w,f)$ for a given open string geometry, and can be computed
directly by localization. For instance, when $G=\bbZ_p$ and denoting chiral
insertions from twisted sectors by $\mathbf{1}_{k}$, $k\in \bbZ_p$, we have
\bea
D^{X,L}\left(d, f\right) &=:& \sum_{k=0}^{p-1}D_k^{X,L}\left(d, f\right)
\mathbf{1}_{k} \nn \\
D_k^{X,L}\left(d, f\right) &=&
\left(\frac{1}{d}\right)^{\mathrm{age}(\mathbf{1}_k)} \frac{1}{\lfloor
  \frac{d}{r_{\rm
      eff}}\rfloor!}\frac{\Gamma\left(d\mu_1+\left\langle\frac{k\alpha_2}{n}\right\rangle
  +\frac{d}{r_{\rm eff}}\right)}{\Gamma\left(d\mu_1-\left\langle\frac{k\alpha_1}{p}\right\rangle +1\right)}
\label{eq:discfunction}
\eea
where we wrote $\alpha_i$ for the
characters of the $\bbZ_p$ action along the $i^{\rm th}$ leg of the vertex.\\

Let us examine  \eqref{eq:masterformulaorbifolds} more closely. According to
this formula, the open topological A-model
on a background with toric branes is controlled by a dual closed theory on the
same background with gravitational descendants turned on. What is more, the
precise relationship is given through a concrete incarnation of a crucial
object in the theory of matrix models, namely the loop insertion operator. The relation between
the open and the closed model has indeed the same structure as the one between connected
correlators and deformed free energies in matrix models \cite{Akemann:1997fd}, and is in complete agreement with the mirror
description in terms of B-branes \cite{Aganagic:2003qj}: the open
topological free energy is obtained by the action of a $1^{\rm st}$-order
differential operator in an infinite number of new modes on a deformed closed
amplitude. In our language, the Ooguri-Vafa operator would take the form 
\bea
e^\LL &:=& \sum_{h=0}^\infty g_s^h \prod_{i=1}^h \LL(w_i,f) \\
F^{\rm open} &=& e^\LL \cF^{\rm closed, desc}\big|_{\rm s.q.c.}
\eea
where we denoted by $\cA\big|_{\rm s.q.c.}$ the reduction to small quantum co-homology.
On the other hand, the insertion of a toric brane - or more precisely, in B-model language,
the insertion of a determinant
%\footnote{See also \cite{Kozcaz:2010af} for a similar  point of view in the context of the AGT conjecture.} 
in
the mirror Kodaira-Spencer theory \cite{Aganagic:2003qj} - can be recast in the
form of a gravitational background shift as
\beq
t_{\a,p}\to t_{\a,p}+\sum_{d=1}^\infty D_\a^{X,L}(d,f) w^d d^{p+1}
\eeq
where $w$ is the A-model open string modulus. \\

It is instructive to look at the particular case $g=0$, $h=1$. When
$g=0$, $h=1$, Eq.~\eqref{eq:masterformulaorbifolds} states that the winding number
$d$ contribution to the disc amplitude takes the compact form
\beq
\oint_{w=0} \frac{1}{2\pi i w^{d+1}}F^{X,L}_{0,1}(t_1, \dots, t_n; w, f) =
D_\a^{X,L}(d,f) J^\a\l(t_1, \dots, t_n; f; \frac{1}{d}\r)
\label{eq:f01}
\eeq
in terms of the $T$-equivariant $J$-function \eqref{eq:Jfun} of $X$.

\subsubsection{General toric Calabi-Yau threefolds}
Up to now we have only considered a particular case of toric Calabi-Yau
threefolds, namely, toric orbifolds of $\bbC^3$. However, we claim that the master formula
\eqref{eq:masterformulaorbifolds} holds true when $X$ is a general TCY3, and
in any patch of its stringy moduli space. \\

\begin{figure}[t]
\centering
\includegraphics[scale=1.2]{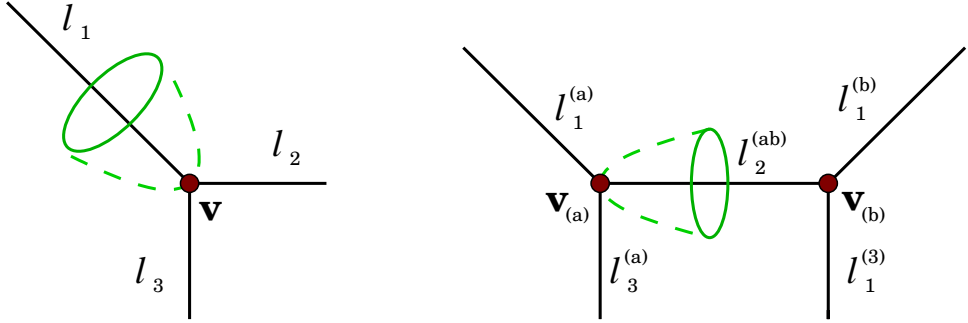}
\caption{\small A toric brane ending on an external leg (left) and an internal leg
  (right) of the web diagram.}
\label{fig:branes}
\end{figure}

A first way to see this, and a more natural one from the point of view of
localization, is that the computation of open string invariants for toric
branes ending on a vertex of the web diagram of $X$ is essentially a local
operation\footnote{We thank Renzo Cavalieri for enlightening discussions of
  this point.}. As explained in \cite{Graber:2001dw}, the bulk geometry affects the
open string amplitude only by replacing \eqref{eq:compact} by a term
corresponding to primary closed string insertions - i.e., for local geometries
given by neighbourhoods of a rigid curve or surface $\Sigma \hookrightarrow X$,
by the push-pull of the normal bundle to $\Sigma$ on
$\overline{\cM}_{g,n}(\Sigma,\beta)$ -, capped with pull-backs of co-homology
classes $\phi_{\a_i}$, $i=1,\dots n$ at the $n$ insertion points. On the
other hand, the local geometry is entirely controlled by the vertex
computation of the previous section, with the (possibly $G$-twisted) chiral
operators of the local theory that are lifted to (possibly $G$-twisted) operators
in the full chiral ring of $X$. Two cases are possible; we restrict here the
discussion to the case in which $X$ is smooth. Suppose first that the leg $l_1$
on which the brane ends is external (see Fig. \ref{fig:branes}), and let $v$ be the tri-valent vertex to which it is
connected; we will call $C_v\simeq\bbC^3$ the affine patch of $X$ associated to
the vertex $v$. Then the loop insertion operator \eqref{eq:lio} for this setup
has the same form with a disc contribution $D_\a^{X,L}(d,f)$ given by
\beq
D_\a^{X,L}(d,f)=D_{\rm id}^{C_V\simeq \bbC^3,L}(d,f)\delta_{\a,\a_V}
\label{eq:discouter}
\eeq
where $\phi_{\a_V}\in H_T(X)$ is the equivariant class of the ``tip''
of the disc attaching at the origin of $C_v$ \cite{Graber:2001dw}. When the
brane intersects an inner leg, we have two fixed vertices $v^{(a)}$ and
$v^{(b)}$  the marked point can
attach to (Fig. \ref{fig:branes}). Then
\beq
D_\a^{X,L}(d,f)=D_{\rm id}^{\bbC^3,L}(d,f)(\delta_{\a,\a_{v^{(a)}}}+\delta_{\a,\a_{v^{(b)}}})
\label{eq:discinner}
\eeq

%1552
\subsubsection{Moving in the open and closed moduli space}
\label{sec:moving}
There is a second vantage point to look at \eqref{eq:masterformulaorbifolds} for a
general pair $(X, L)$, which is motivated by open string mirror
symmetry\footnote{For a different point of view, see also the very recent preprint \cite{renzodusty}.}. The
derivative of the B-model disc amplitude, 
which captures the (infinitesimal)
domain-wall tension of a $D5$ brane wrapping the mirrors of toric branes, should
be a holomorphic, globally defined function of the vector multiplets
\cite{Hori:2000ck, Bouchard:2007ys}.   Let $\hat X$ denote the mirror of $X$, $\cM_{\hat X}$ be the natural toric
compactification of its complex moduli space \cite{MR1677117}, and
$\mathfrak{c}\subset \cM_{\hat X}$ be a chamber in $\cM_{\hat X}$.
Also in view of \cite{Brini:2010sw}, it is
natural to speculate that the B-brane superpotential in a specific chamber should take the form
\beq
\sum_{d=1}^\infty D^{(\mathfrak{c})}(d,f) \cdot I^{(\mathfrak{c})}\l(\mathbf{x}; \frac{1}{d},f\r) y^d
\label{eq:discIfun}
\eeq
where $y$ is the B-model open modulus $y$, $I^{(\mathfrak{c})}(\mathbf{x}; ,f)$ is Givental's
$I$-function in chamber $\mathfrak{c}$ \cite{MR1653024, MR2276766}, and $\mathbf{x}$ are
B-model co-ordinates around the relevant boundary point. The $I$-function is a
co-homology valued generalized hypergeometric series in the variables
$\mathbf{x}$, whose form 
we can read off from the fan $\cF_X$ (see Appendix \ref{sec:Ifunc}),
and whose components
provide a basis of solutions for the $T$-equivariant Picard-Fuchs system
\cite{MR1653024} associated to $X$. The $I$-function is closely related to the
$J$-function \eqref{eq:Jfun} restricted to small quantum
co-homology, as we will see in a moment; we refer the reader to
Appendix \ref{sec:Ifunc} for a detailed account on $J$ and $I$ functions of toric
threefolds. \\

As we emphasized, the disc amplitude is a holomorphic globally defined
quantity; on the other hand, $I$-functions in different chambers should be
related to one another by analytic continuation and a $z$-dependent linear automorphism, as they provide bases of
solutions of the same holonomic system of PDEs in the $\mathbf{x}$-variables, namely, the equivariant Picard-Fuchs
system \cite{MR2700280, MR2510741, MR2486673, Forbes:2006sj}. Then
\beq
I^{(\mathfrak{c}')}\l(\mathbf{x}; z,f\r) = M^{\mathfrak{c}\mathfrak{c}'}(z) I^{(\mathfrak{c})}\l(\mathbf{x}; z,f\r)
\label{eq:imoving}
\eeq
for some invertible matrix $M^{\mathfrak{c}\mathfrak{c}'}(z)$, and imposing
invariance of the disc amplitude we get
\beq
D^{(\mathfrak{c}')}(d,f) =\l[ \l[M^{\mathfrak{c}\mathfrak{c}'}\l(\frac{1}{d}\r)\r]^{-1}\r]^T D^{(\mathfrak{c})}(d,f) 
\label{eq:discmoving}
\eeq
which expresses the change of the brane insertion operator when moving from one
phase to another in the stringy moduli space of $X$. \\

\begin{figure}[t]
\centering
\includegraphics{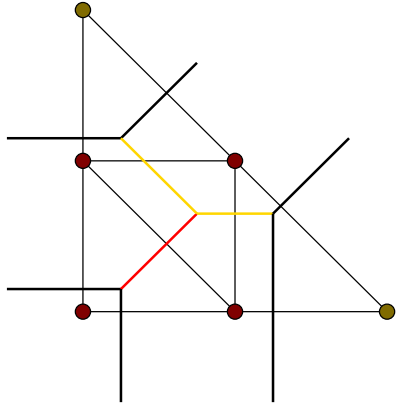}
\caption{\small The minimal $[\bbC^3/G]$ orbifold for which a toric crepant resolution
  contains the resolved conifold $\cO_{\bbP^1}(-1)\oplus \cO_{\bbP^1}(-1)$ as a
  decoupling limit. In this case $G=\bbZ_2 \times \bbZ_2$: one of the large
  radius points corresponds to the closed topological vertex
  geometry, that is, the local CY3 geometry in the neighbourhood of three
  $(-1,-1)$ $\bbP^1$'s touching at a point. In the limit of infinite K\"ahler
  volume for two (gold rays/lines) of the three
  $\bbP^1$, the local geometry reduces to the resolved conifold.}
\label{fig:c3z2z2}
\end{figure}

Formula \eqref{eq:discmoving} and the localization formulae for orbifolds $\bbC^3$ can be used as an alternative to
\eqref{eq:discouter} to compute the brane insertion operator for general toric
Calabi-Yau threefolds. To see this, notice that any TCY3 is a partial crepant
resolution of $\bbC^3/G$ for some $G$, perhaps upon taking the limit of infinite K\"ahler
volume for the curves representing  some the generators of $H_2(X,\beta)$
(Fig.~\ref{fig:c3z2z2}); in particular, in complex dimension three we can
choose $G\simeq \bbZ_p \times \bbZ_q$ for some $p, q\in \bbZ$. 
%\cite{Beasley:1999uz, Feng:2000mi}. 
This can be seen
diagrammatically by adding a finite set $\Theta_X$ of 1-dimensional cones to
the fan $\cF_X$
such that the convex hull of the toric diagram is a triangle; the enlarged
stringy moduli space now incorporates an orbifold point with enhanced $\bbZ_p \times
\bbZ_q$ monodromy. The disc
function in a specific chamber can be computed starting from the one at such
orbifold point, using \eqref{eq:discmoving} to move to any given chamber of
the moduli space of the toric variety associated to the enlarged fan, and
finally decoupling the K\"ahler moduli corresponding to the rays of $\cF_X
\setminus \Theta_X$ (see Figure \ref{fig:c3z2z2}). We will verify explicitly
in the example of Sec. \ref{sec:localp2} the consistency of
\eqref{eq:discmoving} with \eqref{eq:discouter}.  \\

The fact that the B-model superpotential is given by folding the disc
prefactor $D^{(\mathfrak{c})}(d,f)$ with the $I$-function in the relevant
chamber has another interesting consequence. In the local models under
scrutiny, and when restricted to small quantum co-homology \cite{MR2510741, MR2486673}, the generating function of 1-point descendant Gromov-Witten
invariants \eqref{eq:Jfun} and the $I$-function in $A$-model co-ordinates differ by a proportionality
factor\footnote{This is a consequence of the fact that both $J$ and $I$ are
  element of Givental's Lagrangian cone $\LL_X$, that the cone is invariant under
  multiplication by a scalar factor, and by uniqueness properties of the $J$
  function as an element in $\LL_X$. }
\beq
J^{(\mathfrak{c})}\l(\mathbf{t}; z,f\r)= f_{\mathfrak{c}}(\mathbf{t})^{\frac{1}{z}}I^{(\mathfrak{c})}\l(\mathbf{x(t)}; z,f\r)
\label{eq:JvsI}
\eeq
where $f_{\mathfrak{c}}(\mathbf{t})$ is a scalar depending on the closed
string moduli only. Then \eqref{eq:f01} and the invariance condition
\eqref{eq:discIfun} readily imply
that the $A$-model open string flat co-ordinates in chambers $\mathfrak{c}$
and $\mathfrak{c}'$ are related by a renormalization of the type
\beq
w_{\mathfrak{c}}\to \frac{ f_{\mathfrak{c}'}(\mathbf{t})}{
  f_{\mathfrak{c}}(\mathbf{t})} w_{\mathfrak{c}}=   w_{\mathfrak{c}'}
\label{eq:openflat}
\eeq
This is in complete agreement with the expectations from mirror symmetry
\cite{Aganagic:2001nx, Lerche:2001cw, Bouchard:2007ys}, which predict a purely
closed string renormalization of open string moduli. As a consequence, the
master formula \eqref{eq:masterformulaorbifolds} automatically embeds the
change in open string moduli induced by a closed string phase transition! In Sec. \ref{sec:localp2}
we will provide a detailed derivation of this phenomenon for the case of local $\bbP^2$.

\subsection{Equivariant Gromov-Witten invariants and integrable hierarchies}
\label{sec:int}
In the previous section we showed that \eqref{eq:masterformulaorbifolds}
should hold for a general toric Calabi-Yau threefold with toric A-branes, in
all patches of the closed string moduli space, and we discussed in detail what
form the brane insertion operator $\LL(w,f)$ takes in each case. A point we
have not addressed yet is how to compute the second term in
\eqref{eq:masterformulaorbifolds}, namely, the $T$-equivariant free energies
of $X$.  We claim that
\beq
\cF_g(t_{\a,p},f) = \l[\ln \tau(t_{\a,p}+x\delta_{\a,1}\delta_{p,0},f)\r]_{[g]}
\label{eq:flntau}
\eeq
where $\tau(t_{\a,p},f, g_s)$ is the $\tau$-function of a $1+1$
dimensional Hamiltonian integrable hierarchy of evolutionary PDEs
\bea
u^{\a}_{t_{\eta,k}} &= &f_\b^{\a}(u,\eta,k) u_x^\b +
g_s^2 \l[ g_\b^{\a}(u,\eta, k)
u_{xxx}^\b+h_{\b\gamma}^{\a}(u,\eta, k)
u_{xx}^{\b\gamma}+j_{\b\gamma\delta}^{\a}(u,\eta, k)
u_{x}^{\b\gamma\delta}\r]+\cO\l(g_s^4\r) \nn \\ 
&=& \l\{u^{\a}, H_{\eta,k}[u,g_s] \r\} = g_s^2 \frac{\de^2 \log \tau}{\de x \de
  t_{\a,k}}, \qquad k\in \bbN
\label{eq:hierarchy}
\eea
and the subscript $f_{[n]}$ in \eqref{eq:flntau} indicates the $(2n-2)^{\rm nd}$
coefficient in a Taylor-Laurent expansion in $g_s$ around $g_s=0$. In other
words, the string loop expansion corresponds to a gradient expansion in the
$x$-variable, which is in turn identified with the ``puncture'' time variable
$t_{1,0}$. \\

In the rest of the section we will first give a concrete description of the
hierarchy \eqref{eq:hierarchy} at string tree-level, and then discuss how to
deform it to incorporate higher genus corrections. This is for large part a
review of known material, with few extra ingredients to take into account the
gauged $T$-action. The reader is referred to \cite{Witten:1990hr, Dubrovin:1992dz,
  Dubrovin:1994hc, dubrovin-2001, rossi-2009, Brini:2010ap} for more details.

\subsubsection{The tree-level hierarchy and genus zero amplitudes}
\label{sec:planar}
Let us start from the leading order in $g_s$ of \eqref{eq:hierarchy},
\beq
u^{\a}_{t_{\eta,k}} =  f_\b^{\a}(u,\eta,k) u_x^\b 
=  \l\{u^{\a}, H_{\eta,k}[u]_{[0]} \r\} = g_s^2\frac{\de^2 \log \tau_{[0]}}{\de x \de
  t_{\a,k}}
\label{eq:dhierarchy}
\eeq
We want to associate a dynamical system of this form to the planar,
$T$-equivariant A-model on $X$. The functional space the fields $u^\a(x)$
belong to, the (dispersionless) Poisson structure $\l\{, \r\}$, and the tower
of Hamiltonians $H_{\eta,k}[u]_{[0]}$ are constructed as follows \cite{Witten:1990hr, Dubrovin:1992dz,
  Dubrovin:1994hc}:
\ben
\item the phase space $\mathscr{F} \ni u^{\a}(x)$ is given by the loop space
  $L(H_T(X))=C^\infty(S^1, QH_T(X))$ of the
  $T$-equivariant chiral ring;
\item a Poisson structure $\{,\}$ on $\mathscr{F}$ is induced via an $m$-component
  generalization of the KdV poisson bracket as
\beq
\l\{u^\a(x), u^\b(y) \r\}=\eta^{\a\b} \delta'(x-y), \quad \a,\b=1,\dots, m=\mathrm{dim}_{\bbC(\lambda)}QH_T(X)
\label{eq:poisson}
\eeq
where $\eta^{\a\b}$ are the coefficients of the inverse of the $tt$-metric on
$QH_T(X)$, that is, the (possibly orbifold) Poincar\'e pairing on $X$;
\item the tree-level Hamiltonians $H_{\a,p}[u]_{[0]}$ are local densities
\beq
H_{\a,p}[u]_{[0]} = \int_{S^1} h_{\a,p}(u(x))_{[0]} \rd x
\eeq
where $h^{\a}(u,z)_{[0]}=\eta^{\a\b}h_{\a}(u,z)_{[0]}$ are a system of flat co-ordinates
for Dubrovin's deformed Gauss-Manin connection on $H_T(X)$
\beq
\label{eq:nablaz}
\nabla_z = d + \frac{1}{z} \Gamma, \qquad \Gamma_{\a\b}^\gamma= \eta^{\gamma\delta}
\frac{\de^3 F_0}{\de t^\a \de t^\b \de t^\delta}
\eeq
and $h_{\a}(u,z)_{[0]}=: \sum_{p\geq 0} h_{\a,p}(u)_{[0]} z^{-p}$.
\een
The resulting dynamical system is completely specified by the chiral two-point
function $\eta$ and the A-model Yukawa couplings $\Gamma$, both of which
are determined by the toric data
defining $X^{\circlearrowleft T}$: the $tt$-metric can be computed through a degree zero equivariant
co-homology computation on $X$, whereas the Yukawas can be extracted from the
large $z$-expansion of the $J$-function of $X$, which is in turn entirely determined
by the GLSM charge vectors (see Appendix \ref{sec:Ifunc}).
%
%inverse holomorphic 2-point
%function $\eta^{\a\b}$ and by the A-model Yukawa couplings, both of which are
%completely determined by the geometry of
%$X$ and by the weights of the $T$-action: this is trivial to see for the
%former, and is a consequence of the knowledge of the $T$-equivariant J-function
%for the latter (see Appendix \ref{sec:Ifunc}). 
It can then be proven \cite{Dubrovin:1992dz,
  Dubrovin:1994hc} that
\ben
\item the resulting Hamiltonians are in involution
\beq
\l\{H_{\a,p}[u]_{[0]}, H_{\b,q}[u]_{[0]}\r\} = 0 \quad \hbox{for all } \a,\b,p,q 
\eeq
For generic framing, when $T$ acts with compact fixed loci, the family of
Hamiltonian conservation laws determined by $H_{\a,p}[u]_{[0]}$ is complete;
\item the Hamiltonian flows \eqref{eq:dhierarchy} satisfy the $\tau$-symmetry
  condition
\beq
\frac{\de h_{\a,p}(u)_{[0]}}{\de t_{\b,q}} 
= \frac{\de h_{\b,q}(u)_{[0]}}{\de
      t_{\a,p}} = g_s^2 \de_x\frac{\de^2 \log \tau_{[0]}}{\de t_{\a,p} \de t_{\b,q}}
\label{eq:tausymm}
\eeq
in terms of the planar $\tau$-function $\tau_{[0]}$;
\item the logarithm $\cF_0=g_s^2 \ln \tau_{[0]}$ of the genus zero $\tau$-function satisfies the system of
  PDEs \cite{MR1068086, Dijkgraaf:1990nc}
\beq
\frac{\de^3 \cF_0}{\de t_{\a,p} \de t_{\b,q}\de t_{\gamma,r}} = \frac{\de^2
  \cF_0}{\de t_{\a,p} \de t_{\mu,0}} \eta^{\mu \nu} \frac{\de^3 \cF_0}{\de t_{\nu,0} \de t_{\b,q}\de t_{\gamma,r}}
\label{eq:TRR0}
\eeq
as well as the string equation
\beq
\de_x \cF_0 = \sum_{\a,p} t_{\a,p}\de_{t_{\a,p-1}}\cF_0+\frac{1}{2}\eta_{\a\b}t_{\a,0}t_{\b,0}.
\label{eq:string}
\eeq
\een
We refer the reader to \cite{Dubrovin:1994hc} for further details. 
Constructing the dispersionless hierarchy \eqref{eq:dhierarchy} that governs the
$T$-equivariant A-model at tree level just amounts to find a set of flat
co-ordinates for the deformed connection $\nabla_z=0$.
%, by solving a holonomic system of PDEs.  
The genus zero full-descendent free energy $\cF_0(t_{\a,p},f)$
is the potential of the integrability condition \eqref{eq:tausymm}, associated
to an orbit specified by
\beq
u_\a(x)\Big|_{\stackrel{t_{\b,0}=t^\b}{t_{\b,p}=0 \quad p>0}}=t_\a+x\delta_{\a,1}
\label{eq:cauchy}
\eeq
We will see in Sec. \ref{sec:examples} a few concrete examples of this
procedure. \\

We conclude this section with two remarks. First of all, it is natural to ask
how much of this setting goes through to the ordinary, un-gauged case; in fact,
consistently with the discussion in Sec. \ref{sec:closed}, this case corresponds to
a singular limit in which \eqref{eq:dhierarchy} becomes ill-defined, as the inverse $tt$-metric in
\eqref{eq:poisson} vanishes due to non-compactness of the target
space. Secondarily, notice that knowledge of the $T$-equivariant
J-function of $X$ allows to fully reconstruct the descendent theory, by
employing the genus zero topological recursion relations \eqref{eq:TRR0} and
the string equation \eqref{eq:string}. In particular, the generating
function of two-point gravitational descendants with primary insertions
\beq
\bra\bra
\frac{\phi^\a}{z-\psi}, \frac{\phi^\b}{y-\psi}\ket\ket^{X_T}_{0}(\mathbf{t}):=\frac{1}{z w}\sum_{l,m,n=0}^\infty\sum_{\beta\in
H_2(X,\bbZ)} \bra \cO_{\a, l}, \cO_{\b, m}, \mathbf{t}^n
\ket_{0,2+n,\beta}^{X_T} z^{-l}w^{-m},
\eeq
which by \eqref{eq:lio} and
\eqref{eq:masterformulaorbifolds} is the closed string amplitude controlling
the annulus function on the open side, is computed in terms of the J-function as
\cite{Dubrovin:1994hc}
\beq
\bra\bra
\frac{\phi^\a}{z-\psi}, \frac{\phi^\b}{y-\psi}\ket\ket^{X_T}_{0}(\mathbf{t}) =
\frac{1}{z+w}\l(\de_\mu \bra\bra
\frac{\phi^\a}{z-\psi} \ket\ket^{X_T}_{0}(\mathbf{t}) \de^\mu
\bra\bra \frac{\phi^\b}{w-\psi} \ket\ket^{X_T}_{0}(\mathbf{t})-\eta^{\a\b}\r)
\label{eq:annulus}
\eeq
A word of caution is order to compute the r.h.s. of \eqref{eq:annulus}. To
determine the derivative $\de_\mu J$ of the J function we should know the
expression of the latter in {\it big} quantum co-homology - namely, the
quantum parameter $\mathbf{t}$ should take arbitrary values in $QH_T(X)$. On
the other hand, the
Coates-Givental formulae \eqref{eq:Ibundles}, \eqref{eq:Ifunorb} we will use in our applications will
only provide us with the restriction of the $J$-function of $X_T$ to {\it small}
quantum co-homology. However, since the small quantum co-homology ring
generates multiplicatively the entire chiral ring in the toric case, knowledge of the small
J-function is sufficient to compute all the derivatives $\de_\mu J$ in
\eqref{eq:annulus}: two-pointed invariants with one primary insertion can be
computed from the derivatives of the small $J$-function using
\beq
z \nabla_\a \nabla_\b J = \nabla_{\a \star \b} J
\label{eq:Jflat}
\eeq
where $\a \star \b$ in \eqref{eq:Jflat} denotes the operator product of chiral
observables $\phi_\a$, $\phi_\b$. Higher order amplitudes can be computed similarly from \eqref{eq:TRR0}.

\subsubsection{Quasi-Miura transformations and higher genus corrections}
\label{sec:miura}
Having associated a quasi-linear hierarchy to the planar A-model,  the next
and more difficult task is to incorporate higher derivative (i.e., string
loop) corrections
to \eqref{eq:dhierarchy}. A strategy to perform this task in the context of
asymptotically free topological $\sigma$-models coupled to gravity was proposed
in \cite{dubrovin-2001}, to which we refer the reader for an extensive
discussion. In \cite{dubrovin-2001}, the problem of adding higher genus
corrections is cast in terms of a
$g_s$-dependent redefinition of the fields
\beq
u^\a(x)_{[0]} \to u^\a(x, g_s) = g_s^2 \frac{\de^2 \cF}{\de x \de t_{\a,0}}(x,g_s)
\label{eq:miurafields}
\eeq
which sends the dispersionless hierarchy \eqref{eq:dhierarchy} to its
full-dispersive completion \eqref{eq:hierarchy}. The sought-for change of
dependent variables \eqref{eq:miurafields} should be a {\it rational} (or {\it quasi-}){\it
Miura transformation}: namely, in our $\tau$-symmetric context, it takes the
form
\beq
\frac{\cF_0(u)}{g_s^2} \to \cF[u] = \frac{\cF_0(u)}{g_s^2}+ \cF_1(u,u_x,\dots) +
g_s^2 \cF_2(u,u_x,+\dots) + \cO\l(g_s^4\r)
\label{eq:miuratau}
\eeq 
where the coefficient of $g_s^{n}$, $n\geq 0$  in the right hand side is a degree $n$
rational function of the field derivatives $u^{(j)}$ for $j>0$. \\

The (difficult) task of computing higher genus corrections for $X$ and $T$ can then
be viewed as the problem of determining the explicit form of the quasi-Miura
transformation \eqref{eq:miuratau}. In the ordinary A-model
case\footnote{More precisely: for the non-equivariant Gromov-Witten theory of manifolds with
  semi-simple quantum co-homology and vanishing in odd degrees, assuming a
  suitable form of the Virasoro conjecture \cite{Eguchi:1997jd,
    dubrovin-2001}.}, $\cF_g$ can be computed recursively through a set of
differential constraints given by the Dubrovin-Zhang loop equation
\cite{dubrovin-2001}. The equivariant case is however more complicated, and we
should argue differently, as in \cite{Brini:2010ap}. \\
From a practical point of view, we can use the following computational scheme that allows to compute
\eqref{eq:miuratau} at low genus and readily compare with
\cite{Bouchard:2007ys}. The basic idea, as in \cite{Eguchi:2000mi}, is to exploit the existence of universal relations
between co-homology classes on moduli space of stable maps, which highly
constrain the form of the right hand side of \eqref{eq:miuratau}. First of
all, the $3g-2$ theorem of \cite{Eguchi:1998ji, MR2115766} constrains
the dependence on field derivatives of $\cF_g$ to be of the form
\beq
\cF_g[u]= \cF_g\l(u,u_x,u_{xx}, \dots, u^{(3g-2)}\r)
\eeq
Secondarily, higher genus analogues of the topological recursion relations
\eqref{eq:TRR0} exist \cite{Dijkgraaf:1990nc, Kontsevich:1997ex, MR1451505, MR1672112,
  MR717614,
  1998math3072B}. At low genus ($g\leq2$), and combined with the $3g-2$
theorem, they will allow us to recover the
(full-descendant) theory in terms of lower genus gravitational $n$-point
functions and degree zero invariants. For example, at genus 1 we have
\beq
 \frac{\de \cF_1}{\de t_{\a,p}}=\frac{\de^2 \cF_0}{\de t_{\a,p-1} \de
   t_{\nu,0}}\eta^{\mu\nu}\frac{\de \cF_1}{\de t_{\mu,0}}+\frac{1}{24}
 \eta^{\mu \nu}\frac{\de^3 \cF_0}{\de t_{\a,p-1} \de t^\nu \de t^\mu}
\label{eq:TRR1}
\eeq
which determines gravitational $n$-point couplings as a function of the matter
couplings, as well as the tensorial identity on $H_T(X)$
\bea
0 & = & 3 c^\mu_{\a_1\a_2} c^\nu_{\a_3\a_4}  \frac{\de^2 F_1}{\de t_{\mu}\de t^\nu} -4
c^\mu_{\a_1\a_2} c^\nu_{\a_3\mu} \frac{\de^2 F_1}{\de t_{\a_4}\de t^\nu}+ 2
c^\mu_{\a_1 \a_2 \a_3}c^\nu_{\a_4\mu}-c^\mu_{\a_1 \a_2}
c^\nu_{\a_3\a_4\mu}\frac{\de F_1}{\de t^\nu} \nn \\
&+ & \frac{1}{6} c^\mu_{\a_1 \a_2 \a_3}c^\nu_{\a_4\mu\nu}+\frac{1}{24} c^\mu_{\a_1 \a_2 \a_3\a_4}c^\nu_{\mu\nu}-\frac{1}{4} c^\mu_{\a_1 \a_2 \nu}c^\nu_{\a_4\mu\a_3}
\label{eq:g1prim}
\eea
where 
\beq
\bary{ccc}
c^{\mu}_{\a\b}=\eta^{\mu\nu}\frac{\de^3 F_0}{\de t^\a \de t^\b \de t^\nu},
\quad & c^{\mu}_{\a\b\gamma}=\eta^{\mu\nu}\frac{\de^3 F_0}{\de t^\a \de t^\b \de
  t^\gamma\de t^\nu},
\quad & c^{\mu}_{\a\b\gamma\delta}=\eta^{\mu\nu}\frac{\de^3 F_0}{\de t^\a \de
  t^\b \de t^\gamma \de t^\delta\de t^\nu} \\
& F_g(t^\a)=\cF_g(t_{\a,p})\Big|_{\stackrel{t_{\a,p}=0 \hbox{ \footnotesize for }
    p>0}{t_{\a,0}=t^\a}} &
\eary
\eeq
which determines matter couplings in terms of explicitly calculable degree zero
contributions. Analogously, at genus 2 we have two sets of topological
recursion relations. Two identities which will suffice for our purposes are
\bea
\frac{\de \cF_2}{\de t_{\a,p}} &=& \frac{\de^2 \cF_0}{\de t_{\a,p-1}\de
  t_{\b,0}} \frac{\de \cF_2}{\de t_{\gamma,0}}\eta^{\b\gamma}+\frac{\de^2 \cF_0}{\de t_{\a,p-2}\de
  t_{\b,0}} \eta^{\b\gamma}\l(\frac{\de \cF_2}{\de t_{\gamma,1}} - \eta^{\delta\varepsilon}\frac{\de^2 \cF_0}{\de t_{\gamma,0}\de
  t_{\delta,0}}\frac{\de \cF_2}{\de t_{\varepsilon,0}} \r) +\eta^{\varepsilon\gamma} \eta^{\b\delta}\nn \\
& &  \Bigg[ \frac{\de^3 \cF_0}{\de t_{\a,p-2}\de t_{\b,0}\de
  t_{\gamma,0}}\l(\frac{7}{10}
\frac{\de \cF_1}{\de t_{\delta,0}}\frac{\de \cF_1}{\de
  t_{\varepsilon,0}}+\frac{1}{10}\frac{\de^2 \cF_1}{\de t_{\delta,0}\de
  t_{\varepsilon,0}}\r)+\frac{13}{240} \frac{\de^4 \cF_0}{\de t_{\a,p-2}\de t_{\b,0}\de
  t_{\gamma,0}\de t_{\delta,0}}
 \nn \\
& & \frac{\de \cF_1}{\de t_{\varepsilon,0}}-\frac{1}{240} \frac{\de^2 \cF_1}{\de
   t_{\a,p-2}\de t_{\beta,0}}\frac{\de^3 \cF_0}{\de t_{\delta,0}\de
   t_{\gamma,0} \de t_{\varepsilon,0}}+\frac{1}{960} \frac{\de^5 \cF_0}{\de
   t_{\a,p-2}\de t_{\b,0}\de t_{\delta,0}\de
   t_{\gamma,0} \de t_{\varepsilon,0}}
 \Bigg]
\label{eq:TRR2}
\eea
and the Belorusski-Pandharipande equation \cite{1998math3072B} for the matter free energy, which
generalizes \eqref{eq:g1prim} to two-loops. \\

From a conceptual view, the all-genus recursive approach of \cite{dubrovin-2001} was shown by the
authors to be equivalent to Givental's quantization formalism
\cite{MR1901075}; as the latter applies also to non-conformal Frobenius
structures, as long as semi-simplicity of the quantum product is preserved, it
goes through to the case of our interest. This provides in principle a
complete solution to the reconstruction of the higher genus theory; yet, its
concrete implementation is far from trivial. We hope to report on this problem
in detail in the near future. Moreover, in some cases of interest as for example the resolved conifold
\cite{Brini:2010ap, ioguidopaolo} or
configurations of rational curves in a CY3, an all-genus answer can be
obtained through the relation of the planar
hierarchy to known integrable hierarchies, and in particular to known symmetry
reductions of KP/Toda. This gives us the possibility to
use various (sometimes in principle non-perturbative) quantization schemes of the
tree-level hierarchy, which can be used effectively to reconstruct the higher
genus theory \cite{Brini:2010ap}.

\subsubsection{The computational scheme}
Let us summarize concretely how we will apply the machinery of
Sec.~\ref{sec:orbifolds}-\ref{sec:miura} to solve the topological A-model on a
toric open string background $(X,L)$.

\ben
\item  Compute the disc factor $D^{X,L}(d,f)$ that specifies the
  form of the brane insertion operator \eqref{eq:lio}. We can do this in two ways: 
 \bit 
   \item from \eqref{eq:discinner},   \eqref{eq:discouter} using the basic building
     block \eqref{eq:discfunction}, or
   \item using the formula \eqref{eq:discmoving}, and taking the chamber
     $\mathfrak{c}'$ to correspond to a suitable orbifold point of the form $[\bbC^3/G]$
     \eit
 \item Determine the planar hierarchy of $X_T$, where the $T$-action is
    specified by $L$ and the framing as in \eqref{eq:torusaction}. To do that
    we need
   \ben 
     \item the expression of the $tt$-metric $\eta$, and 
     \item the expression for the structure constants $\Gamma$ of $QH^\bullet_T(X)$,
   \een
   both of which are determined by the toric data: $\eta$ by classical
   equivariant intersection theory on $X^{\circlearrowleft T}$, and $\Gamma$
   by the $J$-function using \eqref{eq:Ifun}, \eqref{eq:Ibundles},
   \eqref{eq:Ifunorb}. Multi-pointed amplitudes are computed from the flows
   \eqref{eq:dhierarchy}, or from the $J$-function and the recursion relations
   \eqref{eq:annulus}, \eqref{eq:TRR0}.
\item Compute higher genus corrections by 
  \bit
   \item using the $3g-2$ theorem and universal relations (e.g. \eqref{eq:TRR1},
     \eqref{eq:TRR2}), or
   \item the quantization formalism of semi-simple Frobenius structures \cite{MR1901075}, or
   \item a full-dispersive formulation of the tree-level hierarchy, as in \cite{Brini:2010ap}.
  \eit
\een

We will see in the next section this formalism at work in a number of examples.

\section{Examples}
\label{sec:examples}
\subsection{The framed vertex}
As a first example of our formalism, let us first consider the framed
topological vertex $X=\bbC^3$ with a toric brane ending on the $x_3$
leg. This example was already considered in \cite{kl:open} (see also
\cite{Zhou:2009ea, Zhou:2009gh} for recent work directly relevant to the BKMP theory); what we add here is a
detailed analysis of the dual closed string theory. Eq. \eqref{eq:torusaction} becomes
\beq
\bary{cccc}
(T \simeq \bbC^*) \times  \bbC^3 & \to & \bbC^3\\
(\mu, x_1, x_2,x_3) & \to & (\mu^{f} x_1, \mu^{-f-1}  x_2, \mu x_3)
\eary
\label{eq:Tactvertex}
\eeq
The chiral ring, consisting of the sole identity class, is the trivial algebra
structure on the field of fractions $\bbC(\lambda)$ of $H_T(\mathrm{pt})$
\beq
H_T(X)=\mathrm{span}_{\bbC(\lambda)}\mathbf{1},
\eeq
and the $1\times 1$ $tt$-metric is
\beq
\eta(\mathbf{1}, \mathbf{1})=-\frac{1}{f(f+1)\lambda^3}. 
\label{eq:metricc3}
\eeq
Following the discussion in Sec.~\ref{sec:orbifolds} about $\lambda$-homogeneity of open string
amplitudes, we will henceforth suppress consistently the $\lambda$-dependence
everywhere by setting $\lambda=1$.

\subsubsection{The brane insertion operator}
The brane insertion operator at framing $f$ is simply given by the specialization of
\eqref{eq:discfunction} to the case $p=1$
\beq
D^{\bbC^3, L}= \frac{\Gamma (f d+d)}{d! \Gamma (d f+1)} \mathbf{1}
\label{eq:discfunc3}
\eeq
\subsubsection[The genus zero hierarchy and dKdV]{The genus zero hierarchy and
  dispersionless KdV}
Let us now construct the dispersionless hierarchy governing the
$T$-equivariant tree level theory. Since the chiral ring is a trivial
one-dimensional unital algebra, the deformed Gauss-Manin
connection on $TH_T(X)$ is simply
\beq
\nabla_z = \de_t + \frac{1}{z}
\label{eq:nablazc3}
\eeq
in an affine chart of $H_T(X)$ parameterized by $t\in \bbC(\lambda)$.  Flat
co-ordinates for \eqref{eq:nablazc3} satisfy by definition the ODE
\beq
z \de_t^2 h(t,z) =  \de_t h(t,z).
\label{eq:flatc3eq}
\eeq
A family of solutions of \eqref{eq:flatc3eq} is given by
\beq
h(t,z) = A(z) e^{t/z}+ B(z).
\label{eq:flatc3sol1}
\eeq
The string equation \eqref{eq:string} and comparison with the twisted $J$-function (see Appendix
\ref{sec:Ifunc}) set
\beq
A(z)=z, \qquad B(z)=-z
\eeq
Our sought-for planar hierarchy is then given by a set of compatible
quasi-linear conservation laws on the field space $\mathscr{F}=L(H_T(X))$,
endowed with the Poisson bracket \eqref{eq:poisson}. They take the form
\beq
\frac{\de u(x)}{\de t_{k-1}} := \l\{u(x), \int_{S^1}\frac{u^{k}(y)}{k!}
\r\}= \l\{\bary{cc} 0 & k=0 \\ \frac{u(x)^{k-1}u_x(x)}{(k-1)!} & k>0\eary\r.
\label{eq:dkdv}
\eeq
This hierarchy is the dispersionless limit of the Korteweg-de Vries
hierarchy. The $t^0$ -flow amounts to space translations
\beq
u_{t^0}=u_x
\eeq
whereas the $t^1$-flow 
\beq
u_{t^1}= u u_x
\eeq
is given by the zero-dispersion, $\epsilon\to 0$ limit
of the KdV equation $u_{t^1}= u u_x+\epsilon u_{xxx}$. The orbit corresponding
to the planar full-descendent potential is cut out by the Kontsevich initial datum
\beq
u(t^0=x)\Big|_{t_{k}=0 \hbox{ \footnotesize for } k>0}=x,
\eeq
and the resulting $\tau$-function is therefore the genus zero limit of
the Witten-Kontsevich $\tau$-function \cite{Witten:1990hr,
  Kontsevich:1992ti}. \\

The fact that the closed
planar theory reduces to genus zero topological gravity is expected: the genus
$g$ full-descendent Gromov-Witten potential of $\bbC^3$ is 
\beq
\cF^{\bbC^3,T}_g(t_{p}, f)=\sum_{n=0}^\infty\sum_{p_1,
    \dots, p_n} \frac{\prod_{i=1}^nt_{p_i}}{n!}\bra  \cO_{p_1} \dots  \cO_{p_n}  \ket^{\bbC^3_T}_{g,n}
\label{eq:fulldescc3}
\eeq
where
\beq
\bra  \cO_{p_1} \dots  \cO_{p_n}  \ket^{\bbC^3_T}_{g,n} =
\int_{\cM_{g,n}}\Lambda^\vee_g(1)\Lambda^\vee_g(f)\Lambda^\vee_g(-f-1) \prod_{i=1}^n\psi_i^{p_i}
\label{eq:hodgeone}
\eeq
and $\Lambda^\vee_g(x)=\sum_{i=0}^g(-1)^i x^{g-i} \lambda^{(g)}_i$, where
$\lambda_i^{(g)}=c_i(\mathbb{E}_g)$ is the $i^{\rm th}$ Chern class of the Hodge
bundle on the moduli space of curves; the three Hodge insertions are the
normal contribution of each $\bbC$-fiber to the A-model on the $T$-fixed
point. At genus zero, though, $\Lambda^\vee_0(x)=1$ and and up to a trivial
normalization of the metric \eqref{eq:metricc3} we boil down to the
Witten-Kontsevich case. \\

The resulting integrable structure \eqref{eq:dkdv} is a remarkably
simple one. It is well-known from the topological vertex formalism that
the one-legged framed vertex is governed by a $\tau$-function of the
KP-hierarchy \cite{MR2661523}; the degree of sophistication only increases
when considering two-legged \cite{MR2661523} and three-legged setups
\cite{Aganagic:2003qj}, where the relevant integrable hierarchy coincides
respectively with the $2D$-Toda and the 3-KP hierarchy.  Eq. \eqref{eq:masterformulaorbifolds} gives a
new perspective in terms of simpler 1+1 (as opposed to 2+1) dimensional
integrable systems: the relevant integrable hierachy is a dispersive
deformation of the (simplest) 1+1 dimensional integrable hierarchy, namely the
dKdV hierarchy, and as we will see this statement continues to hold when
considering multi-legged configurations.
%, 2-Toda
%\cite{MR2661523} 3-KP $\tau$-function \cite{Aganagic:2003qj}, \eqref{eq:dkdv} shows
%that it is  in fact (dually) governed by a much simpler system in $1+1$ dimensions (as
%opposed to $2+1$), which at the leading order coincides with dKdV. In fact we
%can say much more: as we will
%show in a moment, moving from one- to multi-legged amplitudes amounts to consider a higher
%dimensional torus action in $\bbC$: this affects our analysis for the
%one-legged case only by trivial framing-dependent normalization factors, and
%in particular it will not affect
%the form of the dKdV flows \eqref{eq:dkdv}.

\subsubsection{String loops and quasi-Miura triviality}

To deform the tree-level hierarchy \eqref{eq:dkdv}, let us apply the machinery
of Sec. \ref{sec:miura}. We will be
looking for a rational Miura transformation of the form
\beq
u(x,g_s) = u(x)_{[0]}+  \frac{\de^2 (g_s^2 \cF([u],g_s)-\cF_0(u))}{\de x^2}
\label{eq:miurauc3}
\eeq
where, using the $3g-2$ theorem, we have
\beq
\cF([u],g_s)- \frac{\cF_0(u)}{g_s^2} =  \cF_1(u_x,u) + g_s^2 \cF_2(u,u_x,u_{xx}, u_{xxx},
u^{(IV)}) + \dots
\label{eq:f-f0c3}
\eeq
In this language the topological recursion relations become a set of
differential identities for the coefficients of the jet variables $u^{(j)}$ in
\eqref{eq:f-f0c3}. For example, at genus 1 \eqref{eq:TRR1} implies
\beq
\cF_1(u,u_x)=\frac{1}{24} \log u_x + F_1(u)
\label{eq:f1c3}
\eeq
where the term $F_1(u)$ is the non-descendent genus one free energy. A trivial
Hodge-integral computation for the only non-vanishing primary invariant at
genus one
 yields
\beq
F_1(u)=\alpha(f) \frac{u}{24}
\eeq
with 
\beq
\a(f)=\frac{f^2+f+1}{f(f+1)}
\eeq
which fixes \eqref{eq:f1c3} completely. \\

At two-loops we can argue similarly and obtain from \eqref{eq:TRR2}
\bea
\cF_2(u,u_{xx},u_{xxx},u^{(IV)}) &=& 
\frac{u^{(4)}(x)}{1152 u'(x)^2}-\frac{7 u^{(3)}(x)
  u''(x)}{1920 u'(x)^3}+\frac{u''(x)^3}{360 u'(x)^4} \nn \\
&+& \frac{7 \a(f) ^2 u''(x)}{5760}-\frac{11 \a(f) 
   u''(x)^2}{5760 u'(x)^2}+\frac{\a(f)  u^{(3)}(x)}{480 u'(x)}+\frac{\b(f)
  u'(x)^2}{5760} \nn \\
\label{eq:f2c3}
\eea
where the parameter $\b(f)$ is fixed by a non-descendent computation on $\overline{\cM}_{2,0}$ as
\beq
\b(f)=-\frac{1}{f(f+1)}
\eeq
Notice that if we set $\a(f)$ and $\b(f)$ equal to zero, we would
be left with the well-known expansion for topological gravity
\cite{Dijkgraaf:1990nc, dubrovin-2001} which sends the dispersionless KdV $\tau$-function to
the all-genus Witten-Kontsevich $\tau$-function. Turning on $\a(f)$ and
$\b(f)$ gives rise to a deformation of the KdV
hierarchy, which we can straightforwardly read off by plugging in
\eqref{eq:miurauc3} into \eqref{eq:dhierarchy}. For the first few flows  we
obtain
\bea
\frac{\de u}{\de t_0} &=& u_x \\
\frac{\de u}{\de t_1} &=&
u(x) u'(x)+\frac{1}{12} g_s ^2 \left[u^{(3)}(x)+\alpha  u'(x) u''(x)\right] +
\frac{g_s^4}{720} \nn \\ & \times & \left[\alpha  \left(u^{(5)}(x)+5 \alpha  u^{(3)}(x) u''(x)\right)+\beta  u^{(3)}(x) u'(x)^2+u'(x) \left(\alpha ^2 u^{(4)}(x)+2 \beta 
   u''(x)^2\right)\right] \nn \\ &+& O\left(g_s ^6\right) \\
\frac{\de u}{\de t_2} &=& \frac{1}{2} u(x)^2 u'(x)+\frac{1}{24} g_s ^2 \left(2
u(x) u^{(3)}(x)+\alpha  u'(x)^3+2 (\alpha  u(x)+2) u'(x)
u''(x)\right)+\frac{g_s^4}{720} \nn \\ & \times &
   \Big[(\alpha  u(x)+3) \left(u^{(5)}(x)+5 \alpha  u^{(3)}(x) u''(x)\right)+u^{(3)}(x) u'(x)^2 \left(\beta  u(x)+5 \alpha ^2\right)+2 \beta  
   \nn \\ & \times &  u'(x)^3 u''(x)+u'(x) \left(\alpha  u^{(4)}(x) (\alpha
   u(x)+8)+2 u''(x)^2 \left(\beta  u(x)+5 \alpha
   ^2\right)\right)\Big]+O\left(g_s ^6\right) \nn \\
\label{eq:hodgekdv}
\eea

The scalar integrable hierarchy that arises, and whose form is apparently new, is
interesting for at least three reasons. First of all it is known that Hodge
integrals can be reduced, using Grothendieck-Riemann-Roch and Faber's
algorithm \cite{MR1728879}, to intersection numbers of
$\psi$ classes on $\overline{\cM}_{g,n}$. From our point of view, \eqref{eq:f1c3} and \eqref{eq:f2c3}
can be regarded as a realization of this statement in the language of
integrable hierarchies: that is, the Korteweg-de Vries hierarchy and its
Hodge-deformation \eqref{eq:hodgekdv} are found to be related by a quasi-Miura
transformation of the form
\bea
u_{\mathrm{KdV}}(x) &=& u_{\mathbb{C}^3}(x) + \frac{\de^2}{\de x^2}
\Bigg[
-\frac{1}{24} g_s ^2 (\alpha  u_{\bbC^3}(x))+\frac{g_s ^4}{5760
  u_{\bbC^3}'(x)^2}\Big(\alpha  u_{\bbC^3}''(x)^2-\beta  u_{\bbC^3}'(x)^4 \nn
\\ &-& 2 \alpha  u_{\bbC^3}^{(3)}(x) u_{\bbC^3}'(x)+3 \alpha ^2 u_{\bbC^3}'(x)^2
   u_{\bbC^3}''(x)\Big)+O\left(g_s ^6\right)
%\frac{1}{24} \alpha  g_s ^2 u_{\mathbb{C}^3}(x)+\frac{g_s ^4}{5760 u_{\mathbb{C}^3}'(x)^2} (-\alpha
%    u_{\mathbb{C}^3}''(x)^2+\beta  u_{\mathbb{C}^3}'(x)^4 \nn \\ &+& 2 \alpha  u_{\mathbb{C}^3}^{(3)}(x) u'(x)+7 \alpha ^2 u_{\mathbb{C}^3}'(x)^2
%   u_{\mathbb{C}^3}''(x))+O\left(g_s ^6\right)
\Bigg]
\eea

Secondarily, it is worthwhile to point out that a remarkable property of the KdV hierarchy goes through to this deformed case: despite the rather involved {\it rational} structure of the Miura
transformation, at the level of the equations of motion we find that the
flows are {\it polynomial} in the jet variables, with a delicate cancellation
of the denominators in the final expressions. It would be interesting to investigate the properties of this
scalar hierarchy in more detail; we plan to investigate this in future
work. \\

Finally, it is interesting to remark that this {\it very same} hierarchy governs the
framed topological vertex with more complicated brane setups, such as the
$3$-legged vertex. Indeed, in this case the dual closed string theory computes
multi-partitions cubic Hodge integrals \cite{Diaconescu:2003qa, Li:2004uf}; at
the level of Eqs. \eqref{eq:masterformulaorbifolds}, \eqref{eq:discfunc3}, 
\eqref{eq:hodgeone}, and \eqref{eq:f1c3}-\eqref{eq:f2c3}, this {\it only amounts} to replace
the framing dependent co-efficients $\alpha(f)$, $\beta(f)$ with $\a(f_1,
\dots, f_l)$, $\b(f_1, \dots, f_l)$, which can be computed exactly as before
starting by replacing \eqref{eq:hodgeone} with
\beq
\int_{\cM_{g,n_1 + n_2 + n_3}}\Lambda^\vee_g(\rho_1)\Lambda^\vee_g(\rho_2)\Lambda^\vee_g(\rho_3) \prod_{i=1}^n\psi_i^{p_i}
\eeq
%
%in general this type of
%scalar hierarchy and to generalize it further to all-genera and higher dimension. 
with $f_i=\rho_{i+1}/\rho_i$; apart from this modification, the underlying integrable hierarchy, will
still have the form \eqref{eq:hodgekdv}.  Computing $l$-legged string amplitudes - that is,
multi-trace mixed correlation functions in a matrix model language - corresponds
to act with loop insertions operator, having the same form
\eqref{eq:discfunc3}, and carrying a different framing parameter
for each leg. For example, for the 2- and 3-legged vertex we find
\beq
\bary{rclrcl}
\a(\rho_1,\rho_2) &=& \frac{\rho_{1}^2+\rho_{1} \rho_{2}+\rho_{2}^2}{\rho_{1}
  \left(\rho_{1} \rho_{2}+\rho_{2}^2\right)} & \b(\rho_1,
\rho_2) &=& -\frac{1}{\rho_1^2 \rho_2 + \rho_1 \rho_2^2} \\
\a(\rho_1,\rho_2, \rho_3) &=& \frac{\rho_{1} \rho_2 + \rho_{2}
  \rho_{3}+\rho_{3} \rho_1}{\rho_{1}\rho_{2}\rho_{3} } & \b(\rho_1,
\rho_2, \rho_3) &=& \frac{1}{\rho_1 \rho_2 \rho_3} \\
\eary
\eeq

\subsubsection{Framed open string amplitudes}
With \eqref{eq:discfunc3} and \eqref{eq:f1c3}, \eqref{eq:f2c3} at hand it is
straightforward to compute open string amplitudes for the framed vertex from
\eqref{eq:masterformulaorbifolds}. Up to the sign ambiguities
affecting open string invariants, we recover the known results for the
topological string on $\bbC^3$, with the coefficients of the $w$-expansion
expressed in closed form in the winding number $d$ and the framing $f$:
\bea
F^{X,L}_{0,1}(f,w) &=& \sum_{d=1}^\infty \frac{\Gamma (f d+d)}{d d!
  \Gamma (d f+1)} w^d , \\
F^{X,L}_{0,2}(f,w_1,w_2) &=&
\sum_{d_1, d_2} \frac{f (f+1) \Gamma (f d_{1}+d_{1}) \Gamma (f
  d_{2}+d_{2})}{(d_{1}-1)! (d_{2}-1)! (d_1+d_2) \Gamma (d_{1} f+1) \Gamma
  (d_{2} f+1)} w_1^{d_1}w_2^{d_2} , \nn \\
\eea
\bea
F^{X,L}_{1,1}(f,w) &=& \sum_{d=1}^\infty \frac{(1-(d-1) f (f+1)) \Gamma (f d+d)}{24 (d-1)! \Gamma (d
  f+1)}w^d, \\
F^{X,L}_{1,2}(f,w_1, w_2) &=& \sum_{d_1, d_2} \frac{\Gamma (f d_{1}+d_{1})
   \Gamma (f d_{2}+d_{2})}{24 d_{1}! d_{2}! \Gamma (d_{1} f+1) \Gamma (d_{2}
  f+1)} w_1^{d_1}w_2^{d_2} \Bigg[d_{1}^2  (-f) (f+1) \nn \\ & \times &+d_{1}
  (1-(d_{2}-1) f (f+1))  - (d_{2}-1) d_{2} f (f+1)+d_{2} \Bigg] , 
\eea
\bea
F^{X,L}_{2,1}(f,w) &=& \sum_{d=1}^\infty \frac {d \Gamma (f d + d)} {5760 (d-1)! \Gamma (d f +      1)}\Bigg[ (5 d^3 f^2 (f + 1)^2 - 
      12 d^2 f (f + 1)  \nn \\ & \times & \left (f^2 + f + 1 \right)+
      7 d \left (f^2 + f + 1 \right)^2 - 
      2 f (f + 1) \Bigg] w^d, \\
F^{X,L}_{2,2}(f,w_1, w_2) &=& \sum_{d_1, d_2} \frac{\Gamma (f d_{1}+d_{1})
  \Gamma (f d_{2}+d_{2}) w_1^{d_1}w_2^{d_2}}{5760 d_{1}! d_{2}! \Gamma (d_{1} f+1) \Gamma
   (d_{2} f+1)} \Bigg[ 5 d_{1}^5 f^2 (f+1)^2+3 d_{1}^4 f \nn \\ & \times &
  (f+1) ((5 d_{2}-4) f (f+1)-4)+d_{1}^3 ((d_{2}-1) f (f+1)-1) \nn \\ & \times
  & ((29 d_{2}-7) f
   (f+1)-7) +  d_{1}^2 \Big(29 d_{2}^3 f^2 (f+1)^2-50 d_{2}^2 f (f+1) \nn \\ & \times
  &
  \left(f^2+f+1\right)+21 d_{2} \left(f^2+f+1\right)^2 - 2 f
   (f+1)\Big)+3 d_{1} d_{2} \nn \\ & \times
  & \Big(5 d_{2}^3 f^2 (f+1)^2-12 d_{2}^2 f (f+1)
  \left(f^2+f+1\right)+ 7 d_{2}  \left(f^2+f+1\right)^2 \nn \\ & -
  & 2 f
   (f+1)\Big)+d_{2}^2 \Big(5 d_{2}^3 f^2 (f+1)^2-12 d_{2}^2 f (f+1) \nn \\ &
  \times & \left(f^2+f+1\right)+7 d_{2} \left(f^2+f+1\right)^2-2 f
   (f+1)\Big)\Bigg]. 
%\dots \nn
\eea

\subsection{The resolved conifold}

As a further example, consider the resolved conifold geometry
$X=\cO_{\bbP^1}(-1)\oplus\cO_{\bbP^1}(-1)$ with a brane on an outer leg. To
simplify our formulae and to make  contact more easily with the results of
\cite{Brini:2010ap}, we will restrict the discussion here to framing one, with the
generalization to arbitrary framing being completely straightforward. \\

\subsubsection{Geometry and phase space data}
\begin{figure}[t]
\centering
\includegraphics{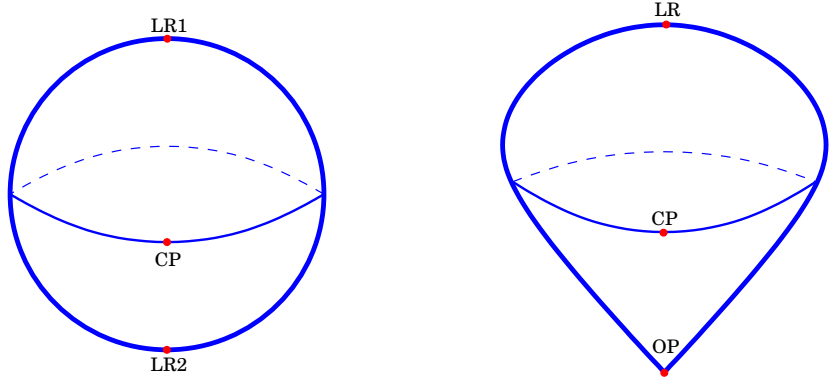}
\caption{\small A pictorial view of the stringy K\"ahler moduli spaces of the resolved conifold
  (left) and local $\bbP^2$ (right). The red dots denote the large radius, orbifold and conifold
  divisors in each case.}
\label{fig:modspaces}
\end{figure}

The toric diagram of $X$ is depicted in Fig. \ref{fig:sigmax}; the skeleton of
its fan is given by the 1-dimensional rays generated by
\beq
v_1 = \l(\bary{c} 0 \\ 0 \\ 1 \eary\r), \quad v_2 = \l(\bary{c} 1 \\ 0 \\ 1 \eary\r), \quad v_3 =
\l(\bary{c} 0 \\ 1\\ 1 \eary\r), \quad v_4 = \l(\bary{c} 1\\ 1 \\ 1\eary\r), \quad
\label{eq:fanconif}
\eeq
We denote by $x_i$ the homogeneous co-ordinate associated to $v_i$. 
The GKZ-extended K\"ahler moduli space $\overline\cM_X$
of $X$ is isomorphic to $\bbP^1$ (Fig. \ref{fig:modspaces}); it has two boundary divisors associated to
the large volume limit, which are related to each other by a toric flop, and a
conifold point. In the ``north'' (resp. ``south'') patch of $\bbP^1$, $X$ is
described as a holomorphic quotient
\beq
X=\frac{\bbC^4 \diagdown Z_X}{(\bbC^*)}
\eeq
where $Z_X=(0,0,x_3,x_4)$ (resp. $Z_X=(x_1,x_2,0,0)$) and we quotient by a $\bbC^*$ action with weights
$(1,1,-1,-1)$. Topological string amplitudes are flop-invariant;
%\cite{Konishi:2006ev}; 
focusing on the ``north'' patch, the $x_1$ and $x_2$
variables are homogeneous co-ordinates for
the null section $X_0 \simeq \bbP^1 \hookrightarrow X$, and $x_3$ and $x_4$
are fiber co-ordinates. The $T$-action in GLSM co-ordinates reads
\beq
\bary{ccccc}
T & \times & X & \to & X \\
\mu, & & (x_1,x_2,x_3,x_4) & \to & (x_1,x_2,\mu x_3, \mu^{-1} x_4) 
\eary
\label{eq:Tactconifold}
\eeq
that is, it covers the trivial action on the base $\bbP^1$ and rotates the
fibers {\it anti}-diagonally. \\

The $T$-equivariant chiral ring localizes classically on the co-homology of
$\bbP^1$. Denote by $\phi_1 := \mathbf{1}=[\bbP^1]^\vee \in H^0(X,\bbC)$, $\phi_2 :=
[\mathrm{pt}]^\vee \in H^2(X, \bbC)$ the identity and the K\"ahler class
respectively. As a vector space, 
\beq
H_T(X) = \mathrm{span}_{\bbC(\lambda)}\{\phi_1, \phi_2 \},
\eeq
endowed with a $tt$-metric given by Atiyah-Bott localization as
\beq
\eta_{ij}= - \delta_{i+j,3} \lambda^2.
\eeq
For $\phi \in H_T(X)$, write $\phi= t^1
\phi_1+t^2\phi_2$. We find from \eqref{eq:fanconif} and \eqref{eq:Ibundles}
that the $J$-function of $X$ with torus action \eqref{eq:Tactconifold} is
\beq
J_X(t^1, t^2, z) = e^{t^1 \phi_1/z+t^2 \phi_2/z }\sum_{d=0}^\infty
\frac{\prod_{-d+1}^0(-\phi_2 + m z +\lambda)(-\phi_2 + m z
  -\lambda)}{\prod_{1}^d(\phi_2 + m z)^2} e^{d t^2} 
\eeq
As before we suppress the $\lambda$ dependence from now on.
\subsubsection{The brane insertion operator}
The brane insertion operator has a particularly simple form in this case. In
the limit of canonical framing, which amounts to switching off the torus
action on the base $\bbP^1$, \eqref{eq:discouter} becomes simply
\beq
D_\a^{X,L}(d)=\frac{1}{d} \delta_{\a,1}
\label{eq:discfunconif}
\eeq

\subsubsection[The genus zero hierarchy and dAL]{The genus zero hierarchy and
  dispersionless Ablowitz-Ladik}

The planar hierarchy has again a very explicit construction, which was
discussed in detail in \cite{Brini:2010ap}. The Christoffel symbols of the
deformed Gauss-Manin connection \eqref{eq:nablaz} can be read off from the large
$z$ asymptotics of $J(z)$
\beq
J_\a(z)=z+ t_\a + \frac{\de_\a F_0}{z}+\cO\l(\frac{1}{z^2}\r)
\eeq
and we find
\beq
\Gamma_{1i}^j = \delta_i^j, \qquad    \Gamma_{22}^1 =
\frac{e^{t^2}}{e^{t^2}-1}, \qquad \Gamma_{22}^2=0
\label{eq:christconif}
\eeq
Flat co-ordinates for $\nabla_z$ satisfy
\beq
z \de_i\de_j h^\a(z) =  \Gamma_{ij}^k\de_k h^\a(z)
\label{eq:flatfuncs}
\eeq
By \eqref{eq:christconif}, the system of PDEs reduces to solve ODEs for
$i=j=1$, $i=1$ and $j=2$, and finally $i=j=2$. 
%By \eqref{eq:discfunconif}, we
%will be interested in finding the deformed flat co-ordinate of the unity, as
%the flows that it generates are the ones that are picked by the brane
%insertion operator. 
Comparison with the
$J$-function of $X$ equivariant w.r.t. \eqref{eq:Tactconifold} (see
\cite{Brini:2010ap} for the complete calculation) yields
\bea
\label{eq:h1}
h^1(t^1,t^2,z) & = & - h_2(t^1, t^2,z) = z\l(e^{t^1/z} \,
  _2F_1\left(-1/z,1/z;1;e^{t^2}\right) -1\r) \\
h^2(t^1,t^2,z) & = & - h_1(t^1, t^2,z) = 
e^{t^1/z}\Bigg[ -1/z \l(\psi ^{(0)} (1+1/z)+\psi ^{(0)}(1-1/z)+2 \gamma\r) \nn \\
& &  _2F_1\left(1/z,-1/z;1;e^{t^2}\right)  - \frac{\pi \left(1-e^{t^2}\right) }{z^2\sin(\pi/z)} \, _2F_1\left(1+1/z,1-1/z;2;1-e^{t^2}\right)
  \Bigg] \nn \\
\label{eq:genintA} 
\eea
where $ \, _2F_1\left(a,b;c;x\right) $ denotes Gauss' hypergeometric function,
$\psi ^{(0)}(z)=\frac{\rd \log \Gamma(z)}{\rd z}$ is the digamma function, and
$\gamma$ is the Euler-Mascheroni constant. \\

As pointed out in \cite{Brini:2010ap}, the resulting hierarchy is the
continuum limit of a 2-component integrable lattice: the Ablowitz-Ladik
hierarchy \cite{MR0377223}. For the first few flows we have
\bea
\frac{\de u^1}{\de t_{1,0}} &=& u^1_x \\
\frac{\de u^2}{\de t_{1,0}} &=& u^2_x \\
\frac{\de u^1}{\de t_{2,0}} &=& - \frac{e^{u^2}}{1-e^{u^2}} u^2_x \\
\frac{\de u^2}{\de t_{2,0}} &=&  u^1_x 
\label{eq:firstflowsconifold}
\eea
By \eqref{eq:discfunconif}, open string amplitudes are controlled in the dual
theory by gravitational descendents of the K\"ahler class\footnote{The fact
  that the open string partition function is insensible to the descendents of
  the unit class is somewhat reminiscent of the $\bbP^1$ topological string,
  where this type of insertions 
%requires a suitable extension of the Toda hierarchy
%  \cite{Carlet:2003pp} and 
are completely invisible both in the mirror symmetry description of
  non-normalizable modes \cite{Aganagic:2003qj} and in its gauge theory realization
  as a deformed $U(1)$ $\cN=2$ Yang-Mills theory in four-dimensions \cite{Marshakov:2006ii}.} $\phi_2$, which are in turn
associated to the $t_{2,p}$ flows of \eqref{eq:h1}. At genus zero, amplitudes
with an arbitrary number of holes  are then completely determined by
\eqref{eq:h1}, \eqref{eq:annulus} and \eqref{eq:TRR0}. We find for example, denoting as
usual $t:=t^2$ the K\"ahler volume of the base $\bbP^1$, 
\bea
F^{X,L}_{0,1}(t,w, f) &=& \sum_{d=1}^\infty  \,
  _2F_1\left(d,-d;1;e^{t}\right) \frac{w^d}{d^2}  \\
F^{X,L}_{0,2}(t,w, f) &=& \sum_{d_1,d_2}
\frac{w_1^{d_1}w_2^{d_2} e^t}{d_{1}+d_{2}} \Bigg[d_{2} \, _2F_1\left(-d_{1},d_{1};1;e^t\right) \,
   _2F_1\left(1-d_{2},d_{2}+1;2;e^t\right) \nn \\ &+& d_{1} \,
   _2F_1\left(1-d_{1},d_{1}+1;2;e^t\right) \,
   _2F_1\left(-d_{2},d_{2};1;e^t\right)\Bigg]
\eea
in perfect agreement with the known results.

\subsubsection{The higher genus theory}
We have two ways to add loop corrections: we could either deform the planar hierarchy 
%and compute the
%first higher genus corrections, 
by using the universal identities
\eqref{eq:TRR1}-\eqref{eq:TRR2} or by exploiting knowledge of the full
dispersive hierarchy \cite{MR0377223}. Either way, we find for example that the generating
function of the rational Miura transformation at one loop reads
\beq
\cF_1^X(u^2, u^1_x, u^2_x) =\frac{1}{24}\log\l( u^1_x(x)^2+\frac{\lambda^2
  e^{u^2(x)}}{1-e^{u^2(x)}}u^2_x(x)^2\r)+\frac{1}{12}\Li_1(e^{u^2(x)})+\frac{u^2(x)}{24}
\eeq
We obtain
\beq
F^{X,L}_{1,1}(t,w) = \sum_{d=1}^\infty \frac{2 d^2 e^t \left(e^t-1\right) \, _2F_1\left(1-d,d+1;2;e^t\right)+\left(3 e^t-1\right) \, _2F_1\left(-d,d;1;e^t\right)}{24 \left(e^t-1\right)}w^d
\eeq
and at two-loops
\bea
F^{X,L}_{2,1}(t,w) &=&\sum_{d=1}^\infty  \Bigg[ \Big(-20 d^4 e^t \left(e^t-1\right)^2+24 d^3
  \left(e^t-1\right)^2 \left(3 e^t-1\right) - d^2 \left(e^t-1\right) \nn \\ &
  \times & \left(e^t \left(63
   e^t+2\right)+7\right)+4 d \left(e^t-1\right) \left(10 e^t+e^{2
     t}+1\right)-24 e^t \left(e^t+1\right)\Big)  \nn \\ &
  \times & \, _2F_1\left(-d,d;1;e^t\right)-4 d
   \left(e^t-1\right) \left(6 d^2 \left(-4 e^t+3 e^{2 t}+1\right)+10 e^t+e^{2 t}+1\right) \nn \\ &
  \times &\, _2F_1\left(1-d,d;1;e^t\right)\Bigg]\frac{w^d}{5760
  \left(e^t-1\right)^3} 
\eea
in complete agreement with the Wilson-loop computation for the unknot in
Chern-Simons theory.
\subsection{Local $\bbP^2$}
\label{sec:localp2}

Up to this point, we have considered Calabi-Yau geometries whose mirrors are
controlled by genus zero spectral curves. We move here to the case of local
surfaces, for which the mirror geometry is encoded in a family of elliptic
curves. The underlying integrable structure is more difficult to describe in
this case; we will see how our formalism goes through. \\

\begin{figure}[t]
\centering
\includegraphics[scale=0.75]{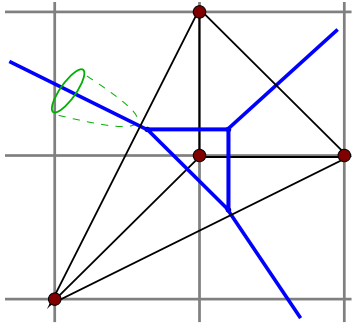}
\caption{\small The toric diagram (black) and the web diagram (blue) of the local $\bbP^2$ geometry with a brane on an
  outer leg.}
\label{fig:localp2}
\end{figure}

As a concrete example, we take $X$ to be the total space $K_{\bbP^2}$ of the canonical line
bundle over the complex projective plane. In order to illustrate the
discussion of Sec.~\ref{sec:moving} in this example, we consider a
configuration given by a toric brane on an outer leg (see
Fig.~\ref{fig:localp2}) at generic framing.

\subsubsection{Geometry and phase space data}
The toric diagram of $X$ is depicted in Fig. \ref{fig:localp2}; the skeleton of
its fan is given by the 1-dimensional rays generated by
\beq
v_1 = \l(\bary{c} 0 \\ 0 \\ 1 \eary\r), \quad v_2 = \l(\bary{c} 1 \\ 0 \\ 1 \eary\r), \quad v_3 =
\l(\bary{c} 0 \\ 1\\ 1 \eary\r), \quad v_4 = \l(\bary{c} -1\\ -1 \\ 1\eary\r), \quad
\eeq
We again denote by $x_i$ the homogeneous co-ordinate associated to $v_i$. The stringy K\"ahler moduli space $\overline\cM_X$
of $X$ is isomorphic to $\bbP^{(1,3)}$ (Fig. \ref{fig:modspaces}); it has one boundary divisor associated to
the large volume limit, a
conifold point, and a $\bbZ_3$-orbifold point, corresponding to the classical
tip of the K\"ahler cone where the $\bbP^2$ divisor shrinks to zero volume. In the large radius patch, $X$ is
described as the holomorphic quotient
\beq
X=\frac{\bbC^{4} \diagdown Z_X}{(\bbC^*)}
\eeq
where now $Z_X=(0,0,0,x_4)$ and we quotient by a $\bbC^*$ action with weights
$(1,1,1,-3)$; $x_i$, $i=1,2,3$ give homogeneous co-ordinates for $\{x_4=0 \} \simeq \bbP^2 \hookrightarrow X$, and $x_4$
co-ordinatizes the $\bbC$-fiber. An ``outer'' brane in this setup intersects
the equator of the fiber; we can take the $T$-action to be
\beq
\bary{ccccc}
T & \times & X & \to & X \\
\mu, & & (x_1,x_2,x_3,x_4) & \to & (x_1,\mu^f x_2,\mu^{-f-1} x_3, \mu x_4) 
\eary
\label{eq:Tactp2}
\eeq
Any permutation of $(x_1,x_2,x_3)$ would yield the same result; this reflects
the symmetry of the three outer legs in Fig. \ref{fig:localp2}. \\

Let $p\in H^2(\bbP^2, \bbC)$ denote the hyperplane class
$c_1(\cO_{\bbP^2}(1))$. We still denote by $p$ its lift to the $T$-equivariant
co-homology of $X$
\beq
H_T(X)=\frac{\bbC(\lambda)[p]}{\bra p^3=\lambda p^2 + f(f+1) \lambda^2 p \ket}
\eeq
 The $tt$-pairing $(\phi_{(1)}, \phi_{(2)})=\sum_{i,j} \eta_{ij} c^i_{(1)} c^j_{(2)}$ between
elements $\phi^{(a)}=\sum_{i=0}^2 c^i_{(a)} p^i$ of the chiral ring is
specified by the Gram matrix in the $p^i$ basis as
\beq
\eta_{ij}= \int_{X^{\circlearrowleft T}} p^{i+j}=\sum_{m=1}^3\Res_{p=p_m}  \frac{p^{i+j}}{p(p+f \lambda)(p-(f+1)\lambda)(-3 p + \lambda)}
\eeq
where $p_1=0$, $p_2=-f \lambda$, $p_3= (f+1)\lambda$. \\
Some of our computations will be more naturally expressed in classical
canonical co-ordinates - that is, in the basis of idempotents $\{\zeta_1,
\zeta_2, \zeta_3\}$ of classical $T$-equivariant co-homology. It is
straightforward to check that 
\bea
\zeta_1 &=& 1 + \frac{1}{f + f^2} \l(\frac{p}{\lambda} - \frac{p^2}{\lambda^2}\r),
\nn \\
\zeta_2 &=& \frac{1}{1 + 3 f + 2 f^2} \l(\frac{f p}{\lambda}+\frac{p^2}{\lambda^2}\r),
\nn \\ 
\zeta_3 &=& \frac{1}{f + 2 f^2} 
\l(\frac{p^2}{\lambda^2}-\frac{(1+f) p}{\lambda}\r)
\label{eq:zeta}
\eea

At the $\bbZ_3$-orbifold point we have $Z_X=(x_1,x_2,x_3,0)$; the resulting
toric variety is the coarse moduli space $\bbC^3/\bbZ_3$ of the orbifold $\cX:=[\bbC^3/\bbZ_3]$,
where the cyclic group acts diagonally with unit weight on $\bbC^3 \simeq
\{(w^{1/3}x_1,w^{1/3}x_2,w^{1/3}x_3)\}$. The $T$-action descends on
$\bbC^3/\bbZ_3$ as
\beq
\bary{clccc}
T & \times & \bbC^3/\bbZ_3 & \to & \bbC^3/\bbZ_3 \\
\mu & , & (w^{1/3}x_1,w^{1/3}x_2,w^{1/3}x_3) & \to &  \quad (\mu^{1/3} w^{1/3}x_1,\mu^{1/3+f}w^{1/3}x_2,\mu^{1/3-f-1}w^{1/3}x_3)
\eary
\eeq
At the orbifold point, the classical chiral ring $H_T^{\rm orb}(\cX)=\bigoplus_{k=1}^3 H_T(\cX_{k})$ is generated by twisted classes
$\mathbf{1}_\frac{k}{3}$, $k=0,1,2$, where $\cX_k$ denotes the $k^{\rm th}$
twisted sector of $\cX$. Write $\psi=\sum_{k=0}^2 d^k \mathbf{1}_\frac{k}{3}$
for $\psi \in H_T^{\rm orb}(\cX)$. The $tt$-metric at this point is just the orbifold
Poincar\'e pairing
\beq
\eta^{\rm orb}_{kl}=- \frac{9 \delta_{k0}\delta_{l0}}{f(f+1) \lambda^3}+\frac{\delta_{k+l,3}}{3}.
\eeq
\subsubsection{The brane insertion operator}
We set $\lambda=1$ as before. We want to compute open topological string
amplitudes for the configuration at hand at the orbifold and at the large
radius point. To do that, we will exploit the point of view of
Sec. \ref{sec:moving}: we use the localization formula at the orbifold point
and impose the invariance condition
\beq
D^{\rm LR}(d,f) \cdot I^{(X)}\l(t; \frac{1}{d},f\r) = D^{\rm orb}(d,f) \cdot
I^{(\cX)}\l(t^{\rm orb}; \frac{1}{d},f\r) 
\label{eq:invariance}
\eeq
where $t$ and $t^{\rm orb}$ are small quantum co-homology parameters. \\

So let us start from the orbifold point. We have from \eqref{eq:discfunction}

\begin{equation}
\label{eq:discfunorb}
D^{\rm orb}\left(d, f\right)=
\frac{1}{\left\lfloor   \frac{d}{3}\right\rfloor!}\left(\frac{1}{d}\right)^{3\langle
  d/3\rangle}
\frac{\Gamma(\frac{d}{3}+\langle
  \frac{d}{3}\rangle+d(f+\frac{1}{3}))}{\Gamma(1-\langle \frac{d}{3}\rangle+d
  (f+\frac{1}{3}))} \mathbf{1}_{3 \bra d/3 \ket}.
\end{equation}
To compute the large radius loop insertion operator we employ
\eqref{eq:invariance} and the chamber-crossing formulas \eqref{eq:imoving},
\eqref{eq:discmoving}. To compute $M^{\rm orb, LR}(z)$, we analytically
continue the large radius $I$-function to the orbifold point, following
\cite{MR2486673}.
%
%
%\section{The brane insertion operator of $K_{\bbP^2}$ from analytic
%  continuation}
%The notation is as in Sec.~\ref{sec:localp2}. Secondary fan. B-model co-ordinates.
%\subsection{The I and J functions at large radius and at the orbifold point}
At the orbifold point, the I-function of $\cX=[\bbC^3/\bbZ_3]$ reads from \eqref{eq:Ifunorb}
\begin{equation}
  \label{eq:IC3Z3}
  I_\cX(x,f, z)  = z \, x^{-1/z}
  \sum_{l \geq 0} {x^l  \over l! \, z^l}
  \prod_{\substack{b : 0 \leq b < {l  \over 3} \\ \bra b\ket = \bra l
        \over 3\ket}} 
  \Big(\frac{1}{3} - b z\Big)   \Big(\frac{1}{3}+f - b z\Big)
  \Big(\frac{1}{3} -f-1- b z\Big)
  \mathbf{1}_{\bra l\over 3\ket}.
\end{equation}
The large $z$ asymptotics fixes
\beq
J_\cX\big(t_{\rm orb},f, z\big) = x^{1/z} I_\cX(x(t_{\rm orb}), f, z) 
\eeq
where
\beq
t_{\rm orb}(x) = \sum_{m \geq 0} (-1)^m {x^{3m+1} \over (3m+1)!}    {\Gamma\big(m+\textstyle{1 \over 3}\big)^3 \over
      \Gamma\big(\textstyle{1 \over 3}\big)^3}.
\eeq

At large radius, we have from \eqref{eq:Ibundles} with $Y=\bbP^2$, $X=K_{\bbP^2}$ and the
torus action \eqref{eq:Tactp2} that
\begin{equation}
  \label{eq:KP2I}
  I_X(y,f, z) = z \sum_{d \geq 0}
  {
    \prod_{-3d < m \leq 0} (1 - 3 p + m z)
    \over \prod_{0 < m \leq d} (p + m z)(p + f+m z)(p -f-1+ m z)} \,
  y^{d + p/z}.
\end{equation}
In this case the large $z$-asymptotics gives for the J-function
\beq
J_X(t,f,z) = e^{f(y)/z} I_X(y,f,z)
\eeq
where
\beq
e^t  = y \exp \big(3 f(y)\big), \qquad  f(y) = \sum_{d>0} \textstyle{(3d-1)! \over (d!)^3} (-y)^d.
\eeq
The power series expansions \eqref{eq:IC3Z3} and \eqref{eq:KP2I} have
respectively radius of convergence $|x|<3$, $|y|<\frac{1}{27}$, where the B-model
variables $x$ and $y$ are local co-ordinates around the orbifold and the large
radius point respectively. Their relation can be read off from the secondary
fan of $X$ (\cite{MR1677117}; see Fig. \ref{fig:modspaces}) to be $y=x^{-3}$. \\

\begin{figure}[hbtp]
\centering
\includegraphics[scale=0.4]{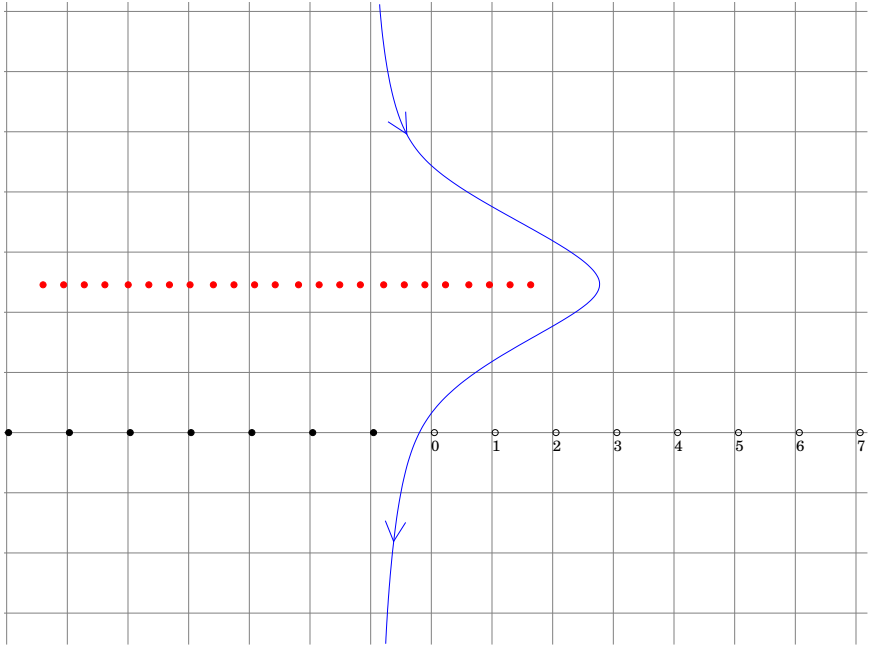}
  \caption{The contour of integration $C$ in \eqref{eq:contourintegral}.}
  \label{fig:contour}
\end{figure}

%\subsection{Analytic continuation}
%\label{sec:ancont}
We follow closely here \cite{MR2486673}, with minor modifications due to the
effectiveness of the torus action on the base of $K_{\bbP^2}$. For the purposes of analytic continuation, it will be worthwhile to rewrite
the summands in \eqref{eq:IC3Z3} and \eqref{eq:KP2I} in terms of ratios of
$\Gamma$-functions
\begin{equation}
  \label{eq:KP2IGamma}
  I_X(y,f, z) = z \sum_{d \geq 0}
  { \Gamma\big(1 + {p \over z} \big)\Gamma\big(1 + {p+f \over z}
    \big)\Gamma\big(1 + {p-f-1 \over z} \big)  \Gamma\big(1 + {1 - 3 p \over z} \big)
    \over \Gamma\big(1 + {p \over z} +d\big)\Gamma\big(1 + {p+f \over z}
    +d \big)\Gamma\big(1 + {p-f-1 \over z} +d \big)
 \Gamma\big(1 + {1 - 3 p \over z} - 3d \big)} \, 
  y^{d + p/z}.
\end{equation}
An efficient way to compute the analytic continuation of $I_X(y,f, z)$ is to use the
Mellin--Barnes method.  We first apply Euler's identity $\Gamma(x) \Gamma(1-x)
= \pi/\sin(\pi x)$ to  \eqref{eq:KP2IGamma}  until each
factor $\Gamma(a + b d)$ in the summand has $b>0$:
\begin{equation}
  \label{eq:IKP2before}
  I_X(y,f, z) = - \Theta_X \sum_{d \geq 0}
  {\Gamma\big(3d-\textstyle{1 - 3p \over z}\big) \over 
    \Gamma\big(1+{p \over z} + d\big)\Gamma\big(1+{p+f \over z} + d\big)\Gamma\big(1+{p-f-1 \over z} + d\big)}
  (-1)^d \, y^{d+p/z}
\end{equation}
with
\beq
\Theta_X = \pi^{-1} z \,
\Gamma\big(1+\textstyle{p \over z}\big) \Gamma\big(1+\textstyle{p+f \over z}\big)\Gamma\big(1+\textstyle{p-f-1 \over z}\big)\,
\Gamma\big(1+\textstyle{1 - 3p \over z}\big) \,
\sin\big(\pi\big[\textstyle{1 - 3p \over z}\big]\big).
\eeq
Consider now the contour integral in the complex $s$-plane as depicted in Figure~\ref{fig:contour}
\begin{equation}
  \label{eq:contourintegral}
  \int_C \Theta_X 
  { 
    \Gamma\big(3s - {1-3p \over z}\big) 
    \Gamma(s) \Gamma(1-s)
    \over
    \Gamma\l(1 + {p \over z} + s\r) \Gamma\l(1 + {p+f \over z} + s\r)\Gamma\l(1 + {p-f-1 \over z} + s\r)
  } y^{s + p/z}.
\end{equation}
The integral \eqref{eq:contourintegral} is defined and
analytic throughout the region $|\arg(y)|<\pi$.  For $|y|<{1 \over
  27}$ we close the contour to the right and pick up the residues at $s=n$,
$n\in\bbZ_+$: this gives us back the large radius expansion \eqref{eq:IKP2before}.  For $|y|>{1 \over 27}$ we close the
contour to the left, and then \eqref{eq:contourintegral} is equal to
the sum of residues at
\begin{align*}
  s = -1-n, & \quad n \geq 0, & \text{and} && s = \textstyle{1 - 3p \over 3z} -
  \textstyle{n \over 3}, & \quad n \geq 0.
\end{align*}
The residues at $s = -1-n$, $n \geq 0$, vanish in $H(Y)$ as they are
divisible by $p^3-p (f + f^2 + p)$.  Thus the analytic continuation $\widetilde{I}_X$
of $I_X$ is equal to the sum of the remaining residues:
\[
\Theta_X \sum_{n \geq 0} {(-1)^n \over 3  n!}
{\Gamma\big(  \textstyle
  { 1 - 3 p \over 3 z} - \textstyle{n \over 3} 
  \big) \Gamma\big(1- \textstyle
  { 1 - 3 p \over 3 z} + \textstyle{n \over 3} 
  \big) \over \Gamma\big(1 + \textstyle {1 \over 3} - {n \over 3}
  \big)\Gamma\big(1 + \textstyle {1 \over 3} - {n \over 3}+\frac{f}{z}
  \big)\Gamma\big(1 + \textstyle {1 \over 3} - {n \over 3}-\frac{f+1}{z}
  \big)} \;
y^{1/3z - n/3}.
\]
Writing this in terms of the co-ordinate $x=y^{-1/3}$, we find that the
analytic continuation $\widetilde{I}_Y(x,z)$ is equal to
\begin{equation}
  \label{eq:IKP2ac}
  {z}\, x^{-1/z} 
  \sum_{n \geq 0} {(-x)^n \over 3.n!}
{\Gamma\big(  \textstyle
  { 1 - 3 p \over 3 z} - \textstyle{n \over 3} 
  \big) \Gamma\big(1- \textstyle
  { 1 - 3 p \over 3 z} + \textstyle{n \over 3} 
  \big) \over \Gamma\big(1 + \textstyle {1 \over 3} - {n \over 3}
  \big)\Gamma\big(1 + \textstyle {1 \over 3} - {n \over 3}+\frac{f}{z}
  \big)\Gamma\big(1 + \textstyle {1 \over 3} - {n \over 3}-\frac{f+1}{z}
  \big)} \;.
\end{equation}
To compute the linear transformation $M:H(X) \to
H(Y)$ that sends the orbifold I-function $I_\cX$ to $\widetilde{I}_X$, it is
sufficient to expand in $x$ the equality
$M(z,f)I_\cX(x,f,z)) = \widetilde{I}_Y(x,f,z)$.
We have
\begin{equation}
  \label{eq:IC3Z3firstfew}
  I_\cX(x,z) = {z}\, x^{-1/z} \l(\mathbf{1}_0 + {x \over z} \mathbf{1}_{1/3} +
  {x^2 \over 2z^2}\mathbf{1}_{2/3} + O(x^3)\r),
\end{equation}
and we obtain from \eqref{eq:IKP2ac} that
\bea
   M (\mathbf{1}_0) &=& 
  {1 \over 3}
  {\sin \big(\pi \big[ {1 - 3p \over z}\big] \big) \over
    \sin \big(\pi \big[ {1 - 3p \over 3z}\big]
    \big)}
  {\Gamma\big(1 + {p \over z}\big)\Gamma\big(1 + {p+f \over z}\big)\Gamma\big(1 + {p-f-1 \over z}\big)  \Gamma\big(1+\textstyle{1 - 3p \over z}\big) \over 
    \Gamma\big(1 + {1 \over 3 z}\big)\Gamma\big(1 + {1 \over 3 z} + \frac{f}{z}\big)\Gamma\big(1 + {1 \over 3 z}-\frac{f+1}{z}\big)} 
  \\
   M(\mathbf{1}_{1/3}) &=& 
  {z \over 3}
  {\sin \big(\pi \big[ {1 - 3p \over z}\big] \big) \over
    \sin \big(\pi \big[ {1 - 3p \over 3z}- {1 \over 3}\big]
    \big)}
  {\Gamma\big(1 + {p \over z}\big)\Gamma\big(1 + {p+f \over z}\big)\Gamma\big(1 + {p-f-1 \over z}\big)   \Gamma\big(1+\textstyle{1 - 3p \over z}\big) \over 
    \Gamma\big(1 + {1 \over 3 z}+ {1 \over 3}\big)\Gamma\big(1 + {1 \over 3 z}+  \frac{f}{z}+{1 \over 3}\big)\Gamma\big(1 + {1 \over 3 z}-\frac{f+1}{z}+ {1 \over 3}\big)} 
 \nn \\ \\
   M(\mathbf{1}_{2/3}) &=& 
  {z^2 \over 3}
  {\sin \big(\pi \big[ {1 - 3p \over z}\big] \big) \over
    \sin \big(\pi \big[ {1 - 3p \over 3z}- {2 \over 3}\big]
    \big)}
  {\Gamma\big(1 + {p \over z}\big)\Gamma\big(1 + {p+f \over
      z}\big)\Gamma\big(1 + {p-f-1 \over z}\big)   \Gamma\big(1+\textstyle{1 - 3p \over z}\big) \over 
    \Gamma\big(1 + {1 \over 3 z}+ {2 \over 3}\big)\Gamma\big(1 + {1 \over 3
      z}+  \frac{f}{z}+ {2 \over 3}\big)\Gamma\big(1 + {1 \over 3
      z}-\frac{f+1}{z}+ {2 \over 3}\big)}. \nn \\
\eea
To determine the change \eqref{eq:discmoving} of the brane insertion operator,
it is convenient to express the images of the isomorphism induced by $M$ in the basis $\zeta_1$,
$\zeta_2$, $\zeta_3$ of the classical idempotents \eqref{eq:zeta} of $H_T(K_{\bbP^2})$. In
this basis, the matrix $M=(m_{ij})$ has the form
\beq
M_{ij}(z)=M(\mathbf{1}_{(j-1)/3})\big|_{p=(1+f)\delta_{i 2}-f \delta_{i 3}}
\eeq
By the degree-twisting condition for the disc function, we have that
\beq
\l[M^{-1}\l(d^{-1}\r)\r]^T \cdot D^{\rm orb}(d,f) = \sum_{i=1}^3
\l[M^{-1}\l(d^{-1}\r)\r]^T_{i, 3 \bra d/3 \ket+1} D^{\rm orb}_{3 \bra d/3
  \ket}(d,f) \zeta_i 
\eeq
For the first few values of $d$ we obtain
\bea
(M^{-1})^T\l(1 \r)_{i, 2} &=& \delta_{i1} \nn \\
(M^{-1})^T\l(1/2 \r)_{i, 3} &=& 2 \delta_{i1} \nn \\
(M^{-1})^T\l(1/3 \r)_{i, 1} &=& -\frac{1}{6} (1+3f)(2+3f)\delta_{i1} \nn \\
(M^{-1})^T\l(1/4 \r)_{i, 2} &=& -\frac{1}{6} (4 f+1) (4 f+3)\delta_{i1}
\eea
which together with \eqref{eq:discfunorb} and \eqref{eq:invariance} yield 
%our
%formulae \eqref{eq:discfunlr} for the
%large radius brane insertion operator.
% the computation is quite lengthy, and the interested reader may find the details in Appendix
%\ref{sec:ancont}. The expression we get for $M^{\rm orb, LR}(z)$ is quite
%complicated, but it simplifies considerably for $1/z=d \in \bbZ$. For the
for the first few values of $d$
\beq
D^{\rm LR}(d,f)= D_1^{\rm LR}(d,f) \zeta_1
\eeq
with
\beq
D_1^{\rm LR}(1,f) = 1, \quad D_1^{\rm LR}(2,f) = - \frac{1 +2 f}{2}, \quad
D_1^{\rm LR}(3,f) =\frac{3}{2} \left(-f-\frac{2}{3}\right)
\left(f+\frac{1}{3}\right), \dots
\label{eq:discfunlr}
\eeq
in complete agreement with \eqref{eq:discouter}. \\

\subsubsection{Open string phase transitions}
As we emphasized in Sec. \ref{sec:moving}, the fact that the I-function is a
globally defined holomorphic function on the stringy moduli space results in a
relative, closed moduli-dependent normalization of the J-functions at the
orbifold and the large radius point. In particular from \eqref{eq:KP2I}, \eqref{eq:IC3Z3} we
have
\bea
J^X(t,z) &=& e^{f(y(t))/(3z)} I^X(y(t),z) \\
J^\cX(t,z) &=& y(t)^{1/(3z)} I^\cX(y(t),z) 
\eea
where 
\beq
f(y)=3 \sum_{d=0}^\infty\frac{(3d-1)!}{(d!)^3}(-y)^d
\eeq
is the worldsheet instanton correction to the closed mirror map. By
\eqref{eq:f01} and \eqref{eq:openflat} this means that the A-model flat {\it
  open} string moduli are related by a renormalization of the form 
\beq
w_{\rm LR} =  Q^{1/3} w_{\rm orb}
\label{eq:openflat2}
\eeq
where $Q$ is the exponentiated K\"ahler parameter; equivalently, in terms of
the B-model open modulus $w_{\rm bare}=e^{-f(y(t))/(3z)} w_{\rm
  LR}=y(t)^{-1/(3z)} w_{\rm orb}$, we have the open mirror maps
\bea
\ln w_{\rm LR} &=& \ln w_{\rm  bare} + \frac{t-\ln y}{3} \\
\ln w_{\rm orb} &=&\ln w_{\rm bare}- \frac{\ln y}{3}
\eea
This is precisely the form of the open mirror map for $K_{\bbP^2}$ and
$[\bbC^3/\bbZ_3]$ that we would obtain from the open string Picard Fuchs system
  \cite{Lerche:2001cw, Aganagic:2001nx, Bouchard:2007ys}, with the correct
  choice of solution automatically picked up at both boundary points.

\subsubsection{Computations}
Having obtained the expression of the brane insertion operators at the
orbifold and the large radius point, we turn to the computation of framed open
string amplitudes in both regions. In each case, we use the expressions
\eqref{eq:KP2I} and \eqref{eq:IC3Z3} for the $T$-equivariant
J-function, and determine recursively higher descendent insertions using
topological recursions relations\footnote{To be precise, as we emphasized in
  Sec. \ref{sec:planar}, formulas such as
  \eqref{eq:annulus} require knowledge of two-point functions with primary
  insertions in big quantum co-homology. However, as $QH_T^\bullet(X)$ is
  generated in degree 2 in the toric case, we can express them in terms of two-point functions
  with one primary insertion in small quantum co-homology, which are in turn
  determined by the small J-function of Appendix \ref{sec:Ifunc}.}. At the orbifold point we have
\bea
\cF^{\rm orb}_{0,1}(t_{\rm orb},w,f) &=&
t_{\rm orb}w -\frac{2f+1}{4} t_{\rm orb}^2 w^2+ \left(\frac{1}{3}+\frac{1}{18} \left(-9 f^2-9 f-2\right)
   t_{\rm orb}^3\right)w^3+\dots \nn \\
\\
\cF^{\rm orb}_{0,2}(t_{\rm orb},w_1,w_2,f) &=&
\frac{9f^2+9f+1}{18}t_{\rm orb}^2w_1 w_2 +\frac{2f+1}{2}\l(w_1^2w_2+w_1
w_2^2\r)+ \dots
\eea
whereas at large radius we find, denoting again with $Q=e^t$ the exponentiated flat
K\"ahler modulus,
\bea
\cF^{\rm LR}_{0,1}(Q,w,f) &=& \left(1-2Q+5 Q^2\right) w-\frac{1}{2}
\left((2 f+1) \left(14 Q^2-4 Q+1\right)\right)w^2+ \dots
%\frac{1}{6} \left(-9 f^2 \left(27 Q^2-6 Q+1\right)-9 f
%   \left(27 Q^2-6 Q+1\right)-54 Q^2+18 Q-2\right)w^3+O\left(w^4\right) 
\\
\cF^{\rm LR}_{0,2}(Q,w_1,w_2,f) &=& \l(-\frac{1}{2} f (f+1)+\left(2 f^2+2
f+1\right) Q+\left(-7 f^2-7 f-4\right) Q^2\r) w_1 w_2 \nn \\ &+&
(1+2f)\l(\frac{1}{3} f (f+1) - \left(2 f^2+2 f+1\right) Q+3  \left(3 f^2+3
f+1\right) Q^2 \r) \nn \\ & & \l(w_1^2
w_2 + w_2 w_1^2\r)+\dots
\eea

It is an expected, yet remarkable fact that framed open string amplitudes with non-trivial
quasi-modular properties are correctly computed using our framework. For
example, the B-model annulus function of local
$\bbP^2$ is linear in the second Eisenstein series $E_2(\tau)$ \cite{Bouchard:2008gu,
  Brini:2008rh}, where $\tau$ is the elliptic modulus of the genus 1 mirror
curve; as a consequence, it transforms non-trivially under changes of duality
frame, and the analytically continued orbifold and large radius amplitudes differ
by a shift. In our language, this shift is correctly reproduced by imposing the
invariance condition for the disc amplitude \eqref{eq:invariance} and the loop
insertion formula \eqref{eq:masterformulaorbifolds}.

\subsection{A genus 3 example}
\begin{figure}[t]
%\begin{minipage}[t]{0.46\linewidth}
\centering
\includegraphics[scale=0.6]{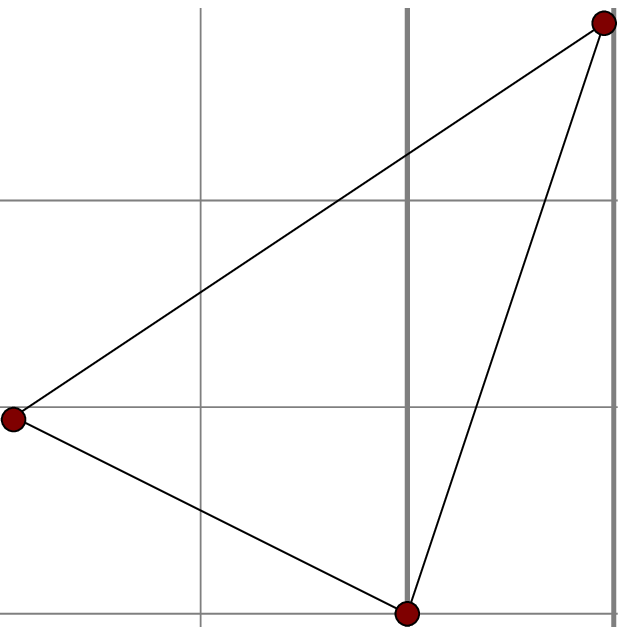}
%\end{minipage} 
%\begin{minipage}[t]{0.08\linewidth}
%\end{minipage}
%\begin{minipage}[t]{0.46\linewidth}
%\centering
%\includegraphics[scale=0.3]{pqwebc3z7.eps}
%\end{minipage} 
\caption{\small The toric diagram 
%(left) and the web diagram (right) of the local Calabi-Yau geometry of three $\bbF_2$ Hirzebruch surfaces glued
%  pairwise along a $\bbP^1$, 
of $[\bbC^3/\bbZ_7]$.}
\label{fig:c3z7}
\end{figure}

As a last example, consider the TCY3 associated to the toric diagram in
Fig. \ref{fig:c3z7}. The resulting singular toric variety $\cX$ is a
$\bbZ_7$-orbifold of $\bbC^3$ by an action with weights $(1,1,5)$;
1-dimensional cones in $\cF_\cX$ can be taken to be
\beq
v_1 = \l(\bary{c} 0 \\ 0 \\ 1 \eary\r), \quad v_2 = \l(\bary{c} -2 \\ 1 \\ 1 \eary\r), \quad v_3 =
\l(\bary{c} 1 \\ 3\\ 1 \eary\r)
\eeq
The (unique) toric crepant resolution of $\cX$ is the local Calabi-Yau geometry of three $\bbF_2$ Hirzebruch surfaces glued
 pairwise along a $\bbP^1$. The mirror geometry is given by a family $\hat X$
 of genus $3$ projective curves over a base $S$ with $\mathrm{dim}_\bbC S=3$;
 it can be shown that  the generic fiber of this family is
 non-hyperelliptic. \\

Computing
open string amplitudes on $\cX$ would be tough using standard methods; first
of all we are away from large radius, and moreover the mirror geometry makes it hard to
compute basic B-model building blocks such as the Bergmann kernel.  In
our formalism, we can use \eqref{eq:masterformulaorbifolds},
\eqref{eq:discfunction}, \eqref{eq:Ifunorb} and \eqref{eq:TRR0}  to address this problem, which is of the same order
of computational difficulty of the $[\bbC^3/\bbZ_3]$ we treated in the last
section. We denote as usual twisted classes by $\mathbf{1}_k$, $k=0,\dots,6$;
we have
\beq
\mathrm{deg}\l(\mathbf{1}_k\r)=2 \mathrm{age}\l(\mathbf{1}_k\r)=\l\{ \bary{cc}
0 & k=0 \\ 2 & 0<k\leq 3 \\ 4 & 3<k\leq 6   \eary  \r.
\eeq
Writing $t_{k}$, $k=1,2,3$ for the (small quantum co-homology) orbifold
K\"ahler parameters, we have, for a toric brane $L$ on one of the two legs acted on
with weight $1$,
\bea
\cF_{0,1}^{\cX,L}(t_1, t_2,t_3,w, f) &=& \l(t_{1}-\frac{7 f+8}{98}  t_{2}
t_3^2\r)w+\l(\frac{t_{2}}{2}-\frac{21 (f+1) t_{1}^2}{294}\r)w^2+\dots \\
\cF_{0,2}^{\cX,L}(t_1, t_2,t_3,w_1,w_2, f) &=& \frac{1}{98} f (f+1) t_{1}^2 w_1
w_2 +\frac{2}{147} f (f+1) t_{2} t_{1}\l(w_1^2 w_2 + w_1 w_2^2 \r)+\dots \nn \\
\eea
\section{Summing the instantons: mirror symmetry and spectral curves}
\label{sec:instantons}
In the last section we checked in detail that our results agree completely with the
computation of A-model topological string amplitudes from string duality; a
natural question to ask is whether we can recover the formalism of
\cite{Aganagic:2003db} or \cite{Bouchard:2007ys} from
\eqref{eq:masterformulaorbifolds}. In this section we begin to address this problem; our aim will be to
make contact with local mirror symmetry and recover the mirror geometry by
resumming A-model instantons at all orders from our localization approach. More precisely, we know that the
reduction $\log y(p)$ of the holomorphic $(3,0)$ form on the Hori-Vafa mirror
curve coincides \cite{Aganagic:2000gs} with
the derivative of the disc amplitude with respect to the B-model open
modulus $p$. In our formalism, this is calculated as in \eqref{eq:discIfun}: once
we have computed the disc factor $D^{X,L}(d,f)$ and the twisted $I$-function
$I^{X}\l(\mathbf{x}; \frac{1}{d},f\r)$, a resummation over winding numbers in
the localization formula will give us the family of spectral curves on the nose
\beq
\log y(p) = \sum_d d D^{X,L}(d,f) I^{X}\l(\frac{1}{d},z,f\r)
\label{eq:resum}
\eeq
We will now see how to recover the mirror Hori-Vafa differentials for the
examples of the previous section.

\subsection{The framed vertex}
\label{sec:vertexsum}
Let us start from the case of the framed vertex; this type of computation was
considered from a different point of view\footnote{Remarkably, in
  \cite{Caporaso:2006gk} the backreaction of the toric brane to K\"ahler
  gravity in this case was found to have a formulation in terms of the A-model
  on non-nef local curves.} in \cite{Aganagic:2001nx, Marino:2001re,
  Caporaso:2006gk, Bouchard:2007ys}. We have from
\eqref{eq:discfunction} that
\beq
\log y(p) = \sum_{d=1}^\infty \frac{\Gamma (f d+d)}{d! \Gamma (d f+1)} p^d
\label{eq:logyc3}
\eeq
We can resum \eqref{eq:logyc3} in hypergeometric form  as
\beq
\log y(p) = \l\{ \bary{lcc}
-\log(1-p) & \mathrm{for} & f=0 \\
p \, _{f+2}F_{f+1}\left(1,1,\frac{f+2}{f+1}, \dots
\frac{2f+1}{f+1};\frac{f+1}{f},\dots, \frac{2f-1}{f},2,2;\frac{(f+1)^{f+1}
   p}{f^f}\right) & \mathrm{for} & f>0
\eary\r. 
\eeq
In order to compare with the topological vertex result, we make the
$GL(2,\bbZ)$ reflection $\log
y(p)\to -\log y(p)$, and we redefine $p \to -p$, which 
amounts to a trivial
shift of the bare open modulus. The resulting sign transformation at each order in $d$
is an ubiquitous fact in open string mirror symmetry, and it should possibly
correspond to the ambiguity in choosing a canonical orientation of the open
string moduli space on the A-model side. We have for the redefined exponentiated
variable $y(p)$ that \cite{Aganagic:2001nx}
\beq
y(p) = \l\{ \bary{lcc}
p+1 & \mathrm{for} & f=0 \\
\frac{f-1}{f} \, _{f}F_{f-1}\left(-\frac{1}{f+1},\frac{1}{f+1}, \dots
\frac{f-1}{f+1};\frac{1}{f},\dots, \frac{f-1}{f};\frac{-(f+1)^{f+1}
   p}{f^{f}}\right) +\frac{1}{f}& \mathrm{for} & f>0
\eary\r.  
\label{eq:speccurvvert}
\eeq
The case of negative $f$  can be recovered from \eqref{eq:speccurvvert} by
using the duality $f \leftrightarrow -f-1$, which is manifest in
\eqref{eq:Tactvertex}. The latter is a hypergeometric root (see {\it e.g.}
\cite{Caporaso:2006gk}) of the trinomial equation
\beq
x y^{-f}-y+1=0
\eeq
which is one form of the spectral curve of the framed vertex, upon identifying
the torus weight $f$ with the
Chern-Simons framing as $f \to -f$.
\subsection{The resolved conifold}
In this case, using \eqref{eq:discfunconif} and \eqref{eq:h1}, the localization formula reads
\beq
\log y(p)= \sum_{d=1}^\infty \, _2F_1\left(-d,d;1;e^t\right)\frac{p^d}{d} 
\eeq
For each fixed integer $d$, the hypergeometric function on the right hand side is a Jacobi polynomial in
$e^t$ \cite{MR2656096}
\beq
 \, _2F_1\left(-d,d;1;e^t\right) = P_n^{0,-1}(1-2 e^t)
\eeq
Using the formula for the generating function of Jacobi polynomials
\beq
\sum_{n=1}^\infty P_n^{\a,\b}(x) z^n = \frac{2^{\alpha +\beta }
 }{R  (R-t+1)^{\alpha } (R+t+1)^{\beta }}-1, \qquad R=\sqrt{1-2 x z + z^2},
\eeq
we obtain
\beq
p \de_p \log{y}(p) = \frac{p+1-\sqrt{\left(4 e^t-2\right) p+p^2+1}}{2 \sqrt{\left(4 e^t-2\right) p+p^2+1}}
\eeq
and integrating once
\beq
\log{y}(p) = -\log \left(\frac{1}{2} \left(\sqrt{4 p e^t+(1-p)^2}-p+1\right)\right)
\eeq
Up to trivial re-definitions $y\to \frac{1}{y}$, $p \to x=-p$, this is just the planar resolvent of the
Chern-Simons matrix model, and we obtain the family of mirror curves of the
resolved conifold in the hyperelliptic,  framing one form
\beq
1+x y +y+ e^t x y^2=0
\eeq 

\subsection{Local surfaces}
The case of toric Calabi-Yau threefolds whose mirror curve has genus greater
than zero presents no extra difficulty (see for example
\cite{Brini:2010sw}). A lengthy, but straightforward general method to compare the sum over instantons
\eqref{eq:resum} with the mirror geometry is to exploit the fact
\cite{Brini:2008rh} that the derivatives of the Hori-Vafa differential with
respect to the B-model closed moduli are algebraic functions of both the open
and the closed moduli. This allows to compare the B-model and A-model
instanton expansion of $\de_{p} \log y(p)$ in a completely explicit way. The
leftover ambiguity is fixed by a computation at the large radius point, which
reduces to the type of sums of Sec. \ref{sec:vertexsum}. \\

We refer the reader
to \cite{Brini:2010sw} for the (rather lengthy) details of this computation
for the case of a Hirzebruch surface $\bbF_2$, and mention here the result for
$K_{\bbP^2}$ with an outer brane at zero framing. Here we find from
\eqref{eq:KP2I} and \eqref{eq:discfunlr} that
\bea
\sum_{d} d w^d D^{X,L}(d,0) I^{X}\l(z, \frac{1}{d},0\r) &=& 
\sum_{d,n} \frac{w^d z^n \Gamma (d (-f-1)+1) (d-1)!}{n!^2 \Gamma (d-3
  n+1) \Gamma (d (-f-1)+n+1)}\Bigg|_{f=0} \nn \\ &=& \log \left(\frac{1}{2}
\left(1-w-\sqrt{1-2 w+ w^2 + 4 w^3 z}\right)\right)\nn \\ &=& \log y(w)
\eea
which yields  the mirror family of elliptic curves of local $\bbP^2$ \cite{Hori:2000ck}
\beq
y^2 + y w+y+z w^3 = 0.
\eeq

\section{Conclusions and outlook}
\label{sec:conclusions}

%\bit
%\item J function nella recoupling section. Dire che J function è completamente
%  determinata dalla geometria. cambiare resummed in closed form negli esempi
%  di c3. spiegare la trivial redefinition nella somma istantoni
%  conifold. controllare il fottuto framing almeno a genere zero per gli esempi
%  di c3. controllare il 1/d in c3 e nel conifold (sul disco era
%  sputtanato). secondary fan e b-model co-ordinates. check framing conifold,
%  dividere per la classe di eulero, open mirror map local $\bbP^2$
%  (+1/2). rifare sistematicamente z <->1/z.
%\item def obs. twisted mass.
%\item worldsheet proof
%\item Givental-Teleman vs ADKMV
%\item SFT vs NS
%\item (refined), backreaction, non-perturbative
%\item multi-legged, inner branes, localization formula away from orbifold point
%\eit

Our formalism opens several new lines of investigation. We mention here a few
ideas for future work.
\bit
\item The most pressing question is whether our approach could open the way
  for a fully rigorous proof of the BKMP proposal
  \cite{Bouchard:2007ys}. In recent work, starting from \cite{Eynard:2009xe},
  it was advocated that one way to do this would be to establish the
  Eynard-Orantin recursion in the form of cut and join equations. Until now,
  this strategy has proved to be fruitful only in the case of genus zero
  spectral curves; yet, the line of reasoning of Sec.~\ref{sec:instantons} lends
  itself to the study of higher genus spectral curves and higher order
  amplitudes, such as the Bergmann kernel. An enticing possibility would be to
    investigate the case of the orbifold topological vertex along the lines of
    \cite{Eynard:2009xe}, and then to give further substance to the arguments
    of Sec.~\ref{sec:moving} that extend the analysis to all chambers of the
    secondary fan.  A  second aspect to understand in our language is the global structure of topological amplitudes; while we checked that our formalism
  embeds automatically the non-trivial quasi-modular shift of the propagator under a change of $S$-duality frame, it would be
  nice to have a further understanding of its origin from the A-model
  side, including, in view of \eqref{eq:masterformulaintro}, the gravitational
  sector.
\item From a physical point of view, it would be desirable to understand
  microscopically the duality \eqref{eq:masterformulaintro}, and perhaps to
  exhibit a worldsheet proof along the lines of \cite{Ooguri:2002gx}. The
  possibility of an interpretation of \eqref{eq:masterformulaintro} from a
  space-time point of view is also an attracting one; on the mirror side, 
   this should find a clear place into the formalism
   \cite{Bonelli:2010cu}. There is plenty to understand here, already for the
   simple example of the resolved conifold. A further aspect to clarify in relation to
   \eqref{eq:masterformulaintro} is the backreaction of toric branes to
   A-model gravity; while \eqref{eq:masterformulaintro} gives a complete
   answer to this problem in terms of gravitational descendents, a more
   satisfactory answer might perhaps be given in a geometric way as in
   \cite{Caporaso:2006gk}. Moreover, the deformation by descendents should
   have a natural interpretation whenever a dual matrix model picture is
   available; for example, in the case of the resolved conifold it is tempting
   to conjecture that the deformation \eqref{eq:discfunconif}, \eqref{eq:h1} should take the
   form of adding higher Casimirs in the dual sum over 2D-partitions \cite{Eynard:2008mt}.
\item While our formalism is, by construction, particularly efficient in
  adding ``holes'' to the worldsheet by repeated application of the brane
  insertion operator $\LL(w,f)$, it is much harder to compute $F_{g,h}^{X,L}$
  at higher genus. As we mentioned, a complete and mathematically rigorous solution is given by Givental's
  quantization formalism; a first line of action would be to find a place for
  a systematic implementation of Givental's formula, and to see how this type of
  quantization relates to the mirror symmetry picture of
  \cite{Aganagic:2003qj}. This is the aspect we are currently devoting more
  attention to, and we hope to report on this in the near future. 
\item Speaking of quantization, the duality with closed
  $T$-equivariant Gromov-Witten invariants might bring into play new
  ingredients: the right hand side of \eqref{eq:masterformulaintro} should
  also be described through the type of
  integrable structures that arise in the study of Gromov-Witten invariants from the
  vantage of Symplectic Field Theory  \cite{Eliashberg:2000fx}. In this
  context,   {\it quantum dispersionless} integrable systems, as opposed to
  the {\it classical dispersionful} hierarchies of Sec.~\ref{sec:int}, appear
  in the description of the full-descendent theory; as
  quantum integrable systems have received much attention recently in the
  study of higher genus corrections to Seiberg-Witten theory \cite{Nekrasov:2009ui}, a possible
  relationship between the two theories begs
  for further understanding.
\item A point which is completely absent in our formalism is the issue of
  Nekrasov's ``refinement'', as we do not have an A-model moduli space interpretation for
  that in terms of holomorphic worldsheet instantons. While it seems natural to conceive that \eqref{eq:masterformulaintro} should
  still hold true in the $\Omega$-background, with the same brane insertion
  operator of the $\beta=1$ case, it is at the moment unclear how to implement
  the deformation in the dual, full-descendent closed string side.
\item Whenever a connection to {\it known} integrable hierarchies is
  available, as for the resolved conifold \cite{Brini:2010ap}, this would
  yield in principle a non-perturbative completion of the open topological
  string: integrable hierarchies arising from KP/Toda reduction as in
  \cite{Brini:2010ap} are naturally well-defined for finite $g_s$,
  (corresponding, {\it e.g.} in the Toda case, to finite lattice spacing), and the analogy with matrix models makes it natural to think that the
  master formula \eqref{eq:masterformulaintro} should hold true
  non-perturbatively in $g_s$. Moreover, the presence of a Lax formalism would
  provide both a new way to derive the mirror geometries (see
  \cite{ioguidopaolo}), as well as a novel method to deform them by
  introducing string loop corrections. 
\item Finally, in the simplest example of a
  topological string related to gauge theory \cite{Marshakov:2006ii}, integrable flows have a natural  interpretation as deforming
   the ultraviolet Lagrangian by turning on descendents of single-trace chiral
  operators. It would be intriguing to generalize this statement in our
  context, and to analyze it in particular for the type of gauge theories that
  the topological string geometrically engineers.

\eit

%\clearpage

\appendix
\section{Toric geometry and I-functions}
\label{sec:Ifunc}
\subsection{The projective case}
We will review here the main statements of \cite{MR1653024, MR2276766,
  MR2510741} about the I-functions for toric orbifolds. To start with, let $X$ be a projective smooth toric variety with $\mathrm{dim}_{\bbC} H^2 (X,
\bbC) = k$, and write $Z_X$ for its Stanley-Reisner ideal. Write
$X$ as the holomorphic quotient 
$X = (\bbC^n \setminus Z)/(\bbC^*)^k$;  the weights
of the torus action can be encoded in an integral $k \times n$ matrix $M = (m_{ij})$. Let $\{C_1,\dots, C_k)$ be a basis of $H_2(X,\bbZ)$ given by
fundamental classes of compact holomorphic curves in $X$ associated to the
rows of $M$, $\{r_1,\dots, r_k\}$
with $r_i\in H^{1,1}(X)$ be their duals in co-homology, and $\{D_1, \dots, D_n\}$ the
divisors given by $z_i=0$, where $z_k$ is the $k^{\rm th}$ homogeneous
co-ordinate of $\bbC^n=(z_1,\dots,z_n)$. We consider furthermore a
$T\simeq(\bbC^*)^r$ multiplicative action on $z_i$; we write $\bbC[\lambda_1, \dots,
  \lambda_r]$ for the coefficient algebra of $H_T(X)$, and write $p_i$ for the
lift of the class $r_i$ to $T$-equivariant co-homology. Consider now the
equivariant classes
\beq
u_j=\sum_{i=1}^n m_{ij}p_i-\lambda_i
\eeq
which are the Poincar\'e duals of the $T$-invariant co-ordinate hyperplanes $D_j$, associated to a 1-dimensional cone of the
secondary fan of $X$; by construction we have $\int_{C_i} u_j = m_{ij}$. The
(small) I-function of $X$  \cite{MR1653024} is defined in terms of the toric data as the
co-homology valued formal power series
\beq
I_X(y_1,\dots, y_k, z)=e^{\ln y_0/z + z \sum_{i=1}^k p_k
  \ln y_k/z}
\sum_{d\in\bbZ_+^k}\prod_{j=1}^n\frac{\prod_{m=-\infty}^0 (u_j+m z)}{\prod_{m=-\infty}^{\sum_i
m_{ij} d^i}(u_j+m z)}\prod_{l=1}^ky_l^{d_l}
\label{eq:Ifun}
\eeq
Let now $J_X(t_1, \dots, t_k, z)$ be the $T$-equivariant J-function of $X$ 
\beq
J_X\l(t_1, \dots, t_k;  z\r) := z+ \sum_{l}t_l p_l +
\sum_{n=0}^\infty\sum_{\beta\in H_2(X,\bbZ)}\bra\mathbf{t}, \dots, \mathbf{t},
\frac{p^l}{z-\psi}\ket^{X_T}_{0,n+1,\beta} p_l
\label{eq:Jfun2}
\eeq
restricted to small
quantum co-homology. Then we have the following

\begin{thm}[Givental] Suppose $c_1^T(X)\geq 0$. Then $J_X=I_X$ up to a
  (weighted homogeneous) change of variables
\bea
\label{eq:ij1}
\ln y_0 & \to & t_0 = \ln y_0+z f_0(y) +h(y), \\
\ln y_l & \to & t_l = \ln y_l+ f_l(y), \quad l=1,\dots, k
\label{eq:ij2}
\eea
where $f_l(y)$, $h(y)$ are analytic functions in $y$.
\end{thm}
\noindent In other words, when $X$ is projective the J-function \eqref{eq:Jfun} is entirely
specified by the toric data defining the I-function, up to the change of variables \eqref{eq:ij1},
\eqref{eq:ij2}. The latter in turn is uniquely determined by comparing the asymptotic
expansions at large $z$ of $J_X$ and $I_X$.
%\beq
%J_X(t_1, \dots, t_k, z) = z + \sum_{l}t_k p_k + \l(\frac{1}{z}\r)
%\eeq

\subsection{The twisted case}
When $X$ is non-compact, and in particular a toric Calabi-Yau threefold,
\eqref{eq:Ifun} continues to hold. The proof of this result will appear in
\cite{ccitnew}, including the case of toric orbifolds. We will state here two
specializations of \cite{ccitnew}, which apply to the cases we treated in
Sec. \ref{sec:examples}, and whose proof has already appeared in the
literature. \\

As a first example, let $X\to Y$ be the total space of a concave line bundle on a
projective semi-positive toric variety $Y$, and let $T'\simeq (\bbC^*)^n$ be a
torus action on $Y$ as in the previous section. We take a  $T\simeq
(\bbC^*)^{n+1} \circlearrowright X$ to cover the $T'$-action on
$Y\hookrightarrow X$ and rotate the fibers by complex multiplication; we write
$\bbC[\lambda_1, \dots, \lambda_n, \lambda_{n+1}]=\bbC[\lambda_1, \dots,
  \lambda_n][\lambda_{n+1}]$ for the coefficient algebra of $H_T^\bullet(X)$,
denoting the equivariant parameter associated to the torus action along the
fibers by $\lambda_{n+1}$. Let
moreover $\rho$ be the first Chern class of
  $X$, and define the following {\it hypergeometric modification}
\beq
  M_X(d) := \prod_{b: \langle \rho,d \rangle < b \leq 0}
  (\lambda_{n+1} + \rho + b z)
\eeq
for $d$ in the semigroup inside $H_2(X,\bbZ)$ generated over $\bbZ_+^k$ by the curve classes $C_1,
\dots, C_k$.
 Then we
have the following \cite{MR2510741, MR2276766}.

\begin{thm}
In terms of the J-function of $Y$ \eqref{eq:Jfun2}, the I-function of
$X$ reads
  \begin{equation}
    \label{eq:Ibundles}
    I_X(t_1, \dots, t_k, z) :=  z \, e^{\sum_{l}p_l \ln y_l /z} 
    \Bigg( 1 + \sum_{d} 
    e^{t_1 d_1} \cdots e^{t_kd_k} \, 
    M_X(d) \,
    \bra p^l \over z(z - \psi)\ket^Y_{0,1,d}
    p_l \Bigg)
  \end{equation}
\end{thm}
 
\subsection{Orbifolds}
The second special case  we need is when $X=[\bbC^3/\bbZ_n]$; the determination of the twisted I-function of
$X$ builds once more on the CCIT-twisting procedure \cite{MR2510741}, this type applied to the
Gromov-Witten theory of $BG$ \cite{MR1950944}. Notations here are as in
Sec.~\ref{sec:orbifolds}.

\begin{thm} \label{thm:BZn}
  Let $X$ be the total space of the direct sum of line bundles $\cE_1 \oplus \cE_2 \oplus \cE_3$ over $B\bbZ_n$.  Let $e_i$
  be the integer such that $\cE_i$ is given by the character $[k]
  \mapsto \exp({2 \pi e_i k \sqrt{-1} \over n})$ of $\bbZ_n$ and that
  $0 \leq e_i < n$.  For $\mathbf{l}=(l_1,\dots, l_n)$, set
\beq
a_i(\mathbf{l})=\sum_{j=1}^n l_j \bra  \frac{ (j-1) e_i}{n}\ket, \quad i=1,2,3
\eeq
Then the small I-function of $X$ reads
\beq
I_X(x_1,\dots, x_n, z) := \sum_{l_1, \dots, l_n} 
  \prod_i \frac{x_i^{l_i}}{l_i! z^{l_i}}  \prod_{j=1}^3 \prod_{m_j=0}^{[a_i(\mathbf{l})]-1}\l(\lambda_j-(\bra a_i(\mathbf{l}) \ket+m_j
    z)\r)  \mathbf{1}_{\bra \sum_i l_i (i-1)/n \ket}
\label{eq:Ifunorb}
\eeq
\end{thm}
When restricted to small quantum co-homology, the I-function
\eqref{eq:Ifunorb} has the large $z$-expansion
\beq
I_X(x_1,\dots, x_n, z) = F(\mathbf{x}) z \mathbf{1}_0+G(\mathbf{x}) + \cO\l(\frac{1}{z}\r)
\eeq
and the J-function is then
\beq
J_X(t_1,\dots, t_n, z) = \frac{I_X(x_1(t),\dots, x_n(t),
  z)}{F(x_1(t), \dots, x_n(t))}
\eeq
where $t_i(\mathbf{x}) = G_i(\mathbf{x})/F(\mathbf{x})$.

\bibliographystyle{plain}
\bibliography{miabiblio}

\end{document}